# Microbial Mat Metagenomes from Waikite Valley, Aotearoa New Zealand

Beatrice Tauer (1), E. Trembath-Reichert (2), LM Ward (1*)

1.      Department of Geosciences, Smith College, Northampton, MA

2.      School of Earth and Space Exploration, Arizona State University, Tempe, AZ

*LWARD15@smith.edu

## Abstract

Objectives:

The rise of complex multicellular ecosystems Neoproterozoic time was preceded by a microbial Proterozoic biosphere, where productivity may have been largely restricted to microbial mats made up of bacteria including oxygenic photosynthetic Cyanobacteria, anoxygenic phototrophs, and heterotrophs. In modern environments, analogous microbial mats can be found in restricted environments such as carbonate tidal flats and terrestrial hot springs. Here, we report metagenomic sequence data from an analog in the hot springs of Waikite Valley, Aotearoa New Zealand, where carbon-rich, slightly-alkaline geothermal waters support diverse phototrophic microbial mats.

Data description:

The Waikite Valley hot spring in the Taupo Volcanic Zone of Aotearoa New Zealand was sampled in duplicate at 8 points along a temperature gradient transect of the outflow, from ~62 °C (near the source) to ~37 °C (~100 meters downstream). ~686 Gb of shotgun metagenomic sequence data was generated by Illumina Novaseq. Each sample was assembled using SPAdes, followed by binning of metagenome-assembled genomes (MAGs) by MetaBAT. These data are useful for the genomic analysis of novel phototrophic bacteria, as well as for ecological comparisons between thermophilic communities with varying temperatures but otherwise similar conditions.

## Objective

The data were collected to investigate the diversity and ecology of microbial mats in organic carbon rich environments. Microbial mats such as these have the potential to act as an analog to Proterozoic microbial mats. Metagenomic sequencing was performed with the assistance of JGI as initial analysis. The data are of interest due to the high abundance and diversity of Chloroflexi, known for their diverse metabolic strategies, including phototrophy and novel pathways for carbon fixation, which appear to have evolved through rampant horizontal gene transfer [1-3]. Detailed analysis of the highest quality MAGs is already underway [4]. Metagenome data will also provide a scaffold for later transcriptomics and other work. The initial binning by JGI produced 1051 high and medium quality MAGs, and although thorough characterization has not yet been performed, it is the focus of ongoing work. We will continue to mine these data for our genomes of interest, but in the meantime, the dataset can serve as a valuable community resource from a novel environment that could lead to the characterization of novel MAGs [e.g. 5-6].

## Data Description

Samples were collected in March 2018 from the Waikite Valley hot spring in the Taupo Volcanic Zone of Aotearoa New Zealand. All samples were collected from microbial mats that were yellow or green in color, leathery, and dense. The microbial mats were submerged in geothermal water from the hot spring with temperatures between 33 and 61.8 °C and a pH of 8 (additional geochemical parameters available through the 1000 Springs Project [7]). Samples measuring ~0.25cm$^3$ were collected along a transect following the natural flow path of the hot spring. Sample collection and processing followed methods described previously [8] and are summarized below.

The collected samples were lysed and DNA preserved immediately after collection in the field utilizing a Zymo Terralyzer BashingBead Matrix and Xpedition Lysis Buffer. Cell lysis was performed with silica beads and a cordless reciprocating saw. Once in the lab, a Zymo Biomics DNA Miniprep Kit was using following manufacturer's instructions to extract and purify microbial DNA. These data include shotgun metagenomic data from samples collected in duplicate from 8 points along a temperature gradient transect of the hot spring, from near the source (~62 °C) to ~100 meters downstream (~37 °C). One pair of samples (Location 6, from a section of the stream at 40 °C) underwent two rounds of DNA extraction from the bead tube in order to test for sequencing consistency.

The Joint Genome Institute via 2x 151bp Illumina NovaSeq was used to perform library preparation and shotgun metagenomic sequencing, generating ~686 Gb of raw data (Table 1, Data set 1). There are 18 samples in total with read counts of 186,232,206 to 340,615,774, total lengths of 521,029,775bp to 1,832,559,633bp, contigs of 456,513 to 1,822,501, and N50s of 22,910 to 131,428. The raw sequence reads were assembled with SPAdes [9] for each metagenomic sample, producing assemblies ranging from 414 MB to ~1 GB in size (Table 1, Data set 2). Metagenome-assembled genomes were binned using MetaBAT v 0.32.4 [10], resulting in 1051 MAGs (Table 1, Data set 3) determined to be medium or high quality. Quality scores were determined with CheckM [11] and are reported according to the MIMAG standards [12].

These data are anticipated to be useful for determining the microbial diversity of this ecosystem, particularly of novel phototrophic bacteria, as well as for ecological comparisons of diversity between thermophilic communities with varying temperatures but otherwise similar conditions.

**Table 1**: Overview of data sets. [13]

| Label | Name of data file/data set | File types (file extension) | Data repository and identifier (DOI or accession number) |
|---|---|---|---|
| Data set 1 | Raw Data | Raw sequence data (.fastq.gz) | JGI GOLD (10.46936/10.25585/60001324) |
| Data set 2 | Assembled contigs | Assembled sequence contigs (.fasta) | JGI GOLD (10.46936/10.25585/60001324) |
| Data set 3 | Metagenome-Assembled Genomes | Genome sequences (.fasta) | JGI GOLD (10.46936/10.25585/60001324) |

Data set 1 : Raw data.

| Sequencing Project Name | File Name |
|---|---|
| LMW ETR NZ 2018 Waikite 1 1 MG | 52494.2.360304.GCACTCAT- GCACTCAT.fastq.gz |
| LMW ETR NZ 2018 Waikite 1 2 MG | 52494.2.360304.CAGGTATC- CAGGTATC.fastq.gz |
| LMW ETR NZ 2018 Waikite 2 1 MG | 52508.4.363682.AACTTGCC- AACTTGCC.fastq.gz |
| LMW ETR NZ 2018 Waikite 2 2 MG | 52508.4.363682.CCAGGATA–CCAGGATA.fastq.gz |
| LMW ETR NZ 2018 Waikite 3 1 MG | 52494.2.360304.TTACCGAG- TTACCGAG.fastq.gz |
| LMW ETR NZ 2018 Waikite 3 2 MG | 52494.2.360304.TTACGGCT- TTACGGCT.fastq.gz |
| LMW ETR NZ 2018 Waikite 4 1 MG | 52494.2.360304.TTCAGGAG- TTCAGGAG.fastq.gz |
| LMW ETR NZ 2018 Waikite 4 2 MG | 52494.2.360304.AATGCCTC- AATGCCTC.fastq.gz |
| LMW ETR NZ 2018 Waikite 5 1 MG | 52508.4.363682.CACCTGTT–CACCTGTT.fastq.gz |
| LMW ETR NZ 2018 Waikite 5 2 MG | 52494.2.360304.GGATTCGT- GGATTCGT.fastq.gz |
| LMW ETR NZ 2018 Waikite 6 1a MG | 52494.2.360304.ACTCAGAC- ACTCAGAC.fastq.gz |
| LMW ETR NZ 2018 Waikite 6 1b MG | 52508.4.363682.CGTCAATG- CGTCAATG.fastq.gz |
| LMW ETR NZ 2018 Waikite 6 2a MG | 52494.2.360304.ATGTAGCG- ATGTAGCG.fastq.gz |
| LMW ETR NZ 2018 Waikite 6 2b MG | 52494.2.360304.GCAAGATC- GCAAGATC.fastq.gz |
| LMW ETR NZ 2018 Waikite 7 1 MG | 52515.4.365442.GTAGAGCA- GTAGAGCA.fastq.gz |
| LMW ETR NZ 2018 Waikite 7 2 MG | 52523.3.366279.CAGCGATT–CAGCGATT.fastq.gz |
| LMW ETR NZ 2018 Waikite S 1 MG | 52494.2.360304.TTGACAGG–TTGACAGG.fastq.gz |
| LMW ETR NZ 2018 Waikite S 2 MG | 52494.2.360304.CAAGTGCA- CAAGTGCA.fastq.gz |

Data set 2: Assembled contigs.

| Sequencing Project Name | File Name |
|---|---|
| LMW ETR NZ 2018 Waikite 1 1 MG | LMW_ETR_NZ_2018__8_FD///assembly.contigs.fasta |
| LMW ETR NZ 2018 Waikite 1 2 MG | LMW_ETR_NZ_2018__5_FD///assembly.contigs.fasta |
| LMW ETR NZ 2018 Waikite 2 1 MG | LMW_ETR_NZ_2018__4_FD///assembly.contigs.fasta |
| LMW ETR NZ 2018 Waikite 2 2 MG | LMW_ETR_NZ_2018__7_FD///assembly.contigs.fasta |
| LMW ETR NZ 2018 Waikite 3 1 MG | LMW_ETR_NZ_2018__6_FD///assembly.contigs.fasta |
| LMW ETR NZ 2018 Waikite 3 2 MG | LMW_ETR_NZ_2018__3_FD///assembly.contigs.fasta |
| LMW ETR NZ 2018 Waikite 4 1 MG | LMW_ETR_NZ_2018__2_FD///assembly.contigs.fasta |
| LMW ETR NZ 2018 Waikite 4 2 MG | LMW_ETR_NZ_2018__FD///assembly.contigs.fasta |
| LMW ETR NZ 2018 Waikite 5 1 MG | LMW_ETR_NZ_2018__16_FD///assembly.contigs.fasta |
| LMW ETR NZ 2018 Waikite 5 2 MG | LMW_ETR_NZ_2018__15_FD///assembly.contigs.fasta |
| LMW ETR NZ 2018 Waikite 6 1a MG | LMW_ETR_NZ_2018__18_FD///assembly.contigs.fasta |
| LMW ETR NZ 2018 Waikite 6 1b MG | LMW_ETR_NZ_2018__17_FD///assembly.contigs.fasta |
| LMW ETR NZ 2018 Waikite 6 2a MG | LMW_ETR_NZ_2018__12_FD///assembly.contigs.fasta |
| LMW ETR NZ 2018 Waikite 6 2b MG | LMW_ETR_NZ_2018__11_FD///assembly.contigs.fasta |
| LMW ETR NZ 2018 Waikite 7 1 MG | LMW_ETR_NZ_2018__14_FD///assembly.contigs.fasta |
| LMW ETR NZ 2018 Waikite 7 2 MG | LMW_ETR_NZ_2018__13_FD///assembly.contigs.fasta |
| LMW ETR NZ 2018 Waikite S 1 MG | LMW_ETR_NZ_2018__10_FD///assembly.contigs.fasta |
| LMW ETR NZ 2018 Waikite S 2 MG | LMW_ETR_NZ_2018__9_FD///assembly.contigs.fasta |

Data set 3: Metagenome-Assembled Genomes.

# Limitations

- The data represent only a single time point collected in 2018, and represent only a small portion of the hot spring system and so may not encompass the full temporal or spatial variability in community composition.
- The MAGs published here are preliminary, and may require additional manual curation prior to use.

# Abbreviations

GOLD: Genomes OnLine Database

JGI: Joint Genome Institute

MAG: metagenome-assembled genome

# Declarations

### Acknowledgments

The authors are indebted to Mark Bowie, the Rotorua Lakes Council, and The Living Waters of Waikite Valley Ltd for providing site access for sampling.

Waikite and the surrounding area was home to the Te Arawa and Ngāti Whakaue iwi of the Māori people for over 500 years prior to seizure by European colonists.

The authors would like to thank the Joint Genome Institute for assistance with sequencing.

# Funding

The work (proposal: 10.46936/10.25585/60001324) conducted by the U.S. Department of Energy Joint Genome Institute (https://ror.org/04xm1d337), a DOE Office of Science User Facility, is supported by the Office of Science of the U.S. Department of Energy operated under Contract No. DE-AC02-05CH11231.

ETR was supported by a NASA Postdoctoral Program fellowship with the NASA Astrobiology Institute.

LMW was supported by the Marine Microbial Ecology Postdoctoral Fellowship from the Simons Foundation and a Geobiology Postdoctoral Fellowship from the Agouron Institute.

# Data availability

The data described in this Data Note can be freely and openly accessed on the JGI GOLD database under GOLD Study ID Gs0151874 [13]. Please see tables 1-3 for details and links to the data.

# Authors' contributions

LMW and ETR conceived of the study, secured funding, and performed field work. LMW processed data. BT and LMW wrote the manuscript. All authors reviewed and edited the manuscript.

# References

1. Ward LM, Shih PM. Granick revisited: Synthesizing evolutionary and ecological evidence for the late origin of bacteriochlorophyll via ghost lineages and horizontal gene transfer. PLoS One 16(1). 2021; https://doi.org/10.1371/journal.pone.0239248

2. Ward LM, Lingappa UF, Grotzinger JP, Fischer WW. Microbial mats in the Turks and Caicos Islands reveal diversity and evolution of phototrophy in the Chloroflexota order Aggregatilineales. Environmental Microbiome 15, 9. 2020; https://doi.org/10.1186/s40793-020-00357-8

3. Shih PM, Ward LM, Fischer WW. Evolution of the 3-hydroxypropionate bicycle and recent transfer of anoxygenic photosynthesis into the Chloroflexi. Proc. Natl. Acad. Sci. U.S.A. 114, 10749–10754. 2017; https://doi.org/10.1073/pnas.1710798114

4. Slosser T, Markert E, Wenick M, Trembath-Reichert E, Ward LM. Metagenome-Assembled Genomes of Nitrososphaera from Aoteroa (New Zealand) Hot Spring Microbial Mats. Research Square, PREPRINT (Version 1). 2023; https://doi.org/10.21203/rs.3.rs-3307453/v1

5. Soo RM, Hemp J, Parks DH, Fischer WW, Hugenholtz P. On the origins of oxygenic photosynthesis and aerobic respiration in Cyanobacteria. Science 355, 1436-1440. 2017; https://doi.org/10.3390/antiox12020423

6. Parks DH, Rinke C, Chuvochina M, Chaumeil P-A, Woodcroft BJ, Evans PN, et al. Recovery of nearly 8,000 metagenome-assembled genomes substantially expands the tree of life. Nat Microbiol 2, 1533–1542. 2017; https://doi.org/10.1038/s41564-017-0012-7

7. Power JF, Carere CR, Lee CK, Wakerley GL, Evans DW, Button M, et al. Microbial biogeography of 925 geothermal springs in New Zealand. Nature communications, 9(1), p.2876. 2018; https://doi.org/10.1038/s41467-018-05020-y

8. Ward LM, Li-Hau F, Kakegawa T, McGlynn SE. Complex history of aerobic respiration and phototrophy in the Chloroflexota class Anaerolineae revealed by high-quality draft genome of Ca. Roseilinea mizusawaensis AA3_104. Microbes and environments, 36(3), p.ME21020. 2021; https://doi.org/10.1264/jsme2.ME21020

9. Prjibelski A, Antipov D, Meleshko D, Lapidus A, Korobeynikov A. Using SPAdes de novo assembler. Current protocols in bioinformatics, 70(1), p.e102. 2020; https://doi.org/10.1002/cpbi.102

10. Kang DD, Froula J, Egan R, Wang Z. MetaBAT, an efficient tool for accurately reconstructing single genomes from complex microbial communities. PeerJ 3: e1165. 2015; https://doi.org/10.7717/peerj.1165

11. Parks DH, Imelfort M, Skennerton CT, Hugenholtz P, Tyson GW. CheckM: assessing the quality of microbial genomes recovered from isolates, single cells, and metagenomes. Genome Res 25: 1043–1055. 2015; https://doi.org/10.1101/gr.186072.114

12. Bowers RM, Kyrpides NC, Stepanauskas R, Harmon-Smith M, Doud D, Reddy TBK, et al. Minimum information about a single amplified genome (MISAG) and a metagenome-assembled genome (MIMAG) of bacteria and archaea. Nat Biotechnol 35: 725–731. 2017; https://doi.org/10.1038/nbt.3893

13. Ward LM, Trembath-Reichert E. Metagenomic characterization of novel phototrophic microbial mats from an organic carbon-rich hot springs. Joint Genome Institute. 2020. https://doi.org/10.46936/10.25585/60001324

Data set 3 : Metagenome-Assembled Genomes.

**Summary table**

| Sequencing Project Name | Files |
|---|---|
| LMW ETR NZ 2018 Waikite 1 1 MG | 87 |
| LMW ETR NZ 2018 Waikite 1 2 MG | 69 |
| LMW ETR NZ 2018 Waikite 2 1 MG | 87 |
| LMW ETR NZ 2018 Waikite 2 2 MG | 83 |
| LMW ETR NZ 2018 Waikite 3 1 MG | 47 |
| LMW ETR NZ 2018 Waikite 3 2 MG | 51 |
| LMW ETR NZ 2018 Waikite 4 1 MG | 86 |
| LMW ETR NZ 2018 Waikite 4 2 MG | 70 |
| LMW ETR NZ 2018 Waikite 5 1 MG | 81 |
| LMW ETR NZ 2018 Waikite 5 2 MG | 48 |
| LMW ETR NZ 2018 Waikite 6 1a MG | 110 |
| LMW ETR NZ 2018 Waikite 6 1b MG | 104 |
| LMW ETR NZ 2018 Waikite 6 2a MG | 116 |
| LMW ETR NZ 2018 Waikite 6 2b MG | 122 |
| LMW ETR NZ 2018 Waikite 7 1 MG | 107 |
| LMW ETR NZ 2018 Waikite 7 2 MG | 88 |
| LMW ETR NZ 2018 Waikite S 1 MG | 101 |
| LMW ETR NZ 2018 Waikite S 2 MG | 119 |
| total | 1576 |

**Table 2**: Data set 3 : Metagenome-Assembled Genomes.

| Sequencing Project Name | File Name |
|---|---|
| LMW ETR NZ 2018 Waikite S 1 MG | 3300044961_10.tar.gz |
| LMW ETR NZ 2018 Waikite S 1 MG | 3300044961_100.tar.gz |
| LMW ETR NZ 2018 Waikite S 1 MG | 3300044961_101.tar.gz |
| LMW ETR NZ 2018 Waikite S 1 MG | 3300044961_102.tar.gz |
| LMW ETR NZ 2018 Waikite S 1 MG | 3300044961_103.tar.gz |
| LMW ETR NZ 2018 Waikite S 1 MG | 3300044961_104.tar.gz |
| LMW ETR NZ 2018 Waikite S 1 MG | 3300044961_105.tar.gz |
| LMW ETR NZ 2018 Waikite S 1 MG | 3300044961_106.tar.gz |
| LMW ETR NZ 2018 Waikite S 1 MG | 3300044961_107.tar.gz |
| LMW ETR NZ 2018 Waikite S 1 MG | 3300044961_108.tar.gz |
| LMW ETR NZ 2018 Waikite S 1 MG | 3300044961_109.tar.gz |
| LMW ETR NZ 2018 Waikite S 1 MG | 3300044961_11.tar.gz |
| LMW ETR NZ 2018 Waikite S 1 MG | 3300044961_110.tar.gz |
| LMW ETR NZ 2018 Waikite S 1 MG | 3300044961_111.tar.gz |
| LMW ETR NZ 2018 Waikite S 1 MG | 3300044961_115.tar.gz |
| LMW ETR NZ 2018 Waikite S 1 MG | 3300044961_117.tar.gz |
| LMW ETR NZ 2018 Waikite S 1 MG | 3300044961_118.tar.gz |
| LMW ETR NZ 2018 Waikite S 1 MG | 3300044961_119.tar.gz |
| LMW ETR NZ 2018 Waikite S 1 MG | 3300044961_12.tar.gz |
| LMW ETR NZ 2018 Waikite S 1 MG | 3300044961_126.tar.gz |
| LMW ETR NZ 2018 Waikite S 1 MG | 3300044961_13.tar.gz |
| LMW ETR NZ 2018 Waikite S 1 MG | 3300044961_131.tar.gz |
| LMW ETR NZ 2018 Waikite S 1 MG | 3300044961_133.tar.gz |
| LMW ETR NZ 2018 Waikite S 1 MG | 3300044961_136.tar.gz |
| LMW ETR NZ 2018 Waikite S 1 MG | 3300044961_137.tar.gz |
| LMW ETR NZ 2018 Waikite S 1 MG | 3300044961_140.tar.gz |
| LMW ETR NZ 2018 Waikite S 1 MG | 3300044961_141.tar.gz |
| LMW ETR NZ 2018 Waikite S 1 MG | 3300044961_142.tar.gz |
| LMW ETR NZ 2018 Waikite S 1 MG | 3300044961_143.tar.gz |
| LMW ETR NZ 2018 Waikite S 1 MG | 3300044961_145.tar.gz |
| LMW ETR NZ 2018 Waikite S 1 MG | 3300044961_147.tar.gz |
| LMW ETR NZ 2018 Waikite S 1 MG | 3300044961_148.tar.gz |
| LMW ETR NZ 2018 Waikite S 1 MG | 3300044961_149.tar.gz |
| LMW ETR NZ 2018 Waikite S 1 MG | 3300044961_150.tar.gz |
| LMW ETR NZ 2018 Waikite S 1 MG | 3300044961_152.tar.gz |
| LMW ETR NZ 2018 Waikite S 1 MG | 3300044961_154.tar.gz |
| LMW ETR NZ 2018 Waikite S 1 MG | 3300044961_156.tar.gz |
| LMW ETR NZ 2018 Waikite S 1 MG | 3300044961_157.tar.gz |
| LMW ETR NZ 2018 Waikite S 1 MG | 3300044961_158.tar.gz |
| LMW ETR NZ 2018 Waikite S 1 MG | 3300044961_163.tar.gz |
| LMW ETR NZ 2018 Waikite S 1 MG | 3300044961_164.tar.gz |
| LMW ETR NZ 2018 Waikite S 1 MG | 3300044961_165.tar.gz |
| LMW ETR NZ 2018 Waikite S 1 MG | 3300044961_166.tar.gz |
| LMW ETR NZ 2018 Waikite S 1 MG | 3300044961_167.tar.gz |
| LMW ETR NZ 2018 Waikite S 1 MG | 3300044961_169.tar.gz |
| LMW ETR NZ 2018 Waikite S 1 MG | 3300044961_170.tar.gz |
| LMW ETR NZ 2018 Waikite S 1 MG | 3300044961_171.tar.gz |
| LMW ETR NZ 2018 Waikite S 1 MG | 3300044961_173.tar.gz |
| LMW ETR NZ 2018 Waikite S 1 MG | 3300044961_174.tar.gz |
| LMW ETR NZ 2018 Waikite S 1 MG | 3300044961_18.tar.gz |
| LMW ETR NZ 2018 Waikite S 1 MG | 3300044961_19.tar.gz |
| LMW ETR NZ 2018 Waikite S 1 MG | 3300044961_21.tar.gz |
| LMW ETR NZ 2018 Waikite S 1 MG | 3300044961_24.tar.gz |
| LMW ETR NZ 2018 Waikite S 1 MG | 3300044961_28.tar.gz |
| LMW ETR NZ 2018 Waikite S 1 MG | 3300044961_29.tar.gz |
| LMW ETR NZ 2018 Waikite S 1 MG | 3300044961_3.tar.gz |
| LMW ETR NZ 2018 Waikite S 1 MG | 3300044961_35.tar.gz |
| LMW ETR NZ 2018 Waikite S 1 MG | 3300044961_38.tar.gz |
| LMW ETR NZ 2018 Waikite S 1 MG | 3300044961_39.tar.gz |
| LMW ETR NZ 2018 Waikite S 1 MG | 3300044961_4.tar.gz |
| LMW ETR NZ 2018 Waikite S 1 MG | 3300044961_44.tar.gz |
| LMW ETR NZ 2018 Waikite S 1 MG | 3300044961_46.tar.gz |
| LMW ETR NZ 2018 Waikite S 1 MG | 3300044961_47.tar.gz |
| LMW ETR NZ 2018 Waikite S 1 MG | 3300044961_48.tar.gz |
| LMW ETR NZ 2018 Waikite S 1 MG | 3300044961_51.tar.gz |
| LMW ETR NZ 2018 Waikite S 1 MG | 3300044961_52.tar.gz |
| LMW ETR NZ 2018 Waikite S 1 MG | 3300044961_53.tar.gz |
| LMW ETR NZ 2018 Waikite S 1 MG | 3300044961_55.tar.gz |
| LMW ETR NZ 2018 Waikite S 1 MG | 3300044961_57.tar.gz |
| LMW ETR NZ 2018 Waikite S 1 MG | 3300044961_59.tar.gz |
| LMW ETR NZ 2018 Waikite S 1 MG | 3300044961_6.tar.gz |
| LMW ETR NZ 2018 Waikite S 1 MG | 3300044961_60.tar.gz |
| LMW ETR NZ 2018 Waikite S 1 MG | 3300044961_61.tar.gz |
| LMW ETR NZ 2018 Waikite S 1 MG | 3300044961_62.tar.gz |

| LMW ETR NZ 2018 Waikite S 1 MG | 3300044961_63.tar.gz |
|---|---|
| LMW ETR NZ 2018 Waikite S 1 MG | 3300044961_67.tar.gz |
| LMW ETR NZ 2018 Waikite S 1 MG | 3300044961_69.tar.gz |
| LMW ETR NZ 2018 Waikite S 1 MG | 3300044961_7.tar.gz |
| LMW ETR NZ 2018 Waikite S 1 MG | 3300044961_70.tar.gz |
| LMW ETR NZ 2018 Waikite S 1 MG | 3300044961_71.tar.gz |
| LMW ETR NZ 2018 Waikite S 1 MG | 3300044961_72.tar.gz |
| LMW ETR NZ 2018 Waikite S 1 MG | 3300044961_73.tar.gz |
| LMW ETR NZ 2018 Waikite S 1 MG | 3300044961_75.tar.gz |
| LMW ETR NZ 2018 Waikite S 1 MG | 3300044961_76.tar.gz |
| LMW ETR NZ 2018 Waikite S 1 MG | 3300044961_77.tar.gz |
| LMW ETR NZ 2018 Waikite S 1 MG | 3300044961_78.tar.gz |
| LMW ETR NZ 2018 Waikite S 1 MG | 3300044961_79.tar.gz |
| LMW ETR NZ 2018 Waikite S 1 MG | 3300044961_80.tar.gz |
| LMW ETR NZ 2018 Waikite S 1 MG | 3300044961_81.tar.gz |
| LMW ETR NZ 2018 Waikite S 1 MG | 3300044961_82.tar.gz |
| LMW ETR NZ 2018 Waikite S 1 MG | 3300044961_85.tar.gz |
| LMW ETR NZ 2018 Waikite S 1 MG | 3300044961_86.tar.gz |
| LMW ETR NZ 2018 Waikite S 1 MG | 3300044961_87.tar.gz |
| LMW ETR NZ 2018 Waikite S 1 MG | 3300044961_89.tar.gz |
| LMW ETR NZ 2018 Waikite S 1 MG | 3300044961_9.tar.gz |
| LMW ETR NZ 2018 Waikite S 1 MG | 3300044961_90.tar.gz |
| LMW ETR NZ 2018 Waikite S 1 MG | 3300044961_91.tar.gz |
| LMW ETR NZ 2018 Waikite S 1 MG | 3300044961_93.tar.gz |
| LMW ETR NZ 2018 Waikite S 1 MG | 3300044961_94.tar.gz |
| LMW ETR NZ 2018 Waikite S 1 MG | 3300044961_95.tar.gz |
| LMW ETR NZ 2018 Waikite S 1 MG | 3300044961_97.tar.gz |

| Sequencing Project Name | File Name |
|---|---|
| LMW ETR NZ 2018 Waikite S 2 MG | 3300044998_1.tar.gz |
| LMW ETR NZ 2018 Waikite S 2 MG | 3300044998_100.tar.gz |
| LMW ETR NZ 2018 Waikite S 2 MG | 3300044998_102.tar.gz |
| LMW ETR NZ 2018 Waikite S 2 MG | 3300044998_103.tar.gz |
| LMW ETR NZ 2018 Waikite S 2 MG | 3300044998_105.tar.gz |
| LMW ETR NZ 2018 Waikite S 2 MG | 3300044998_108.tar.gz |
| LMW ETR NZ 2018 Waikite S 2 MG | 3300044998_110.tar.gz |
| LMW ETR NZ 2018 Waikite S 2 MG | 3300044998_111.tar.gz |
| LMW ETR NZ 2018 Waikite S 2 MG | 3300044998_113.tar.gz |
| LMW ETR NZ 2018 Waikite S 2 MG | 3300044998_114.tar.gz |
| LMW ETR NZ 2018 Waikite S 2 MG | 3300044998_115.tar.gz |
| LMW ETR NZ 2018 Waikite S 2 MG | 3300044998_119.tar.gz |
| LMW ETR NZ 2018 Waikite S 2 MG | 3300044998_122.tar.gz |
| LMW ETR NZ 2018 Waikite S 2 MG | 3300044998_127.tar.gz |
| LMW ETR NZ 2018 Waikite S 2 MG | 3300044998_13.tar.gz |
| LMW ETR NZ 2018 Waikite S 2 MG | 3300044998_130.tar.gz |
| LMW ETR NZ 2018 Waikite S 2 MG | 3300044998_131.tar.gz |
| LMW ETR NZ 2018 Waikite S 2 MG | 3300044998_133.tar.gz |
| LMW ETR NZ 2018 Waikite S 2 MG | 3300044998_134.tar.gz |
| LMW ETR NZ 2018 Waikite S 2 MG | 3300044998_135.tar.gz |
| LMW ETR NZ 2018 Waikite S 2 MG | 3300044998_136.tar.gz |
| LMW ETR NZ 2018 Waikite S 2 MG | 3300044998_139.tar.gz |
| LMW ETR NZ 2018 Waikite S 2 MG | 3300044998_14.tar.gz |
| LMW ETR NZ 2018 Waikite S 2 MG | 3300044998_142.tar.gz |
| LMW ETR NZ 2018 Waikite S 2 MG | 3300044998_144.tar.gz |
| LMW ETR NZ 2018 Waikite S 2 MG | 3300044998_148.tar.gz |
| LMW ETR NZ 2018 Waikite S 2 MG | 3300044998_149.tar.gz |
| LMW ETR NZ 2018 Waikite S 2 MG | 3300044998_15.tar.gz |
| LMW ETR NZ 2018 Waikite S 2 MG | 3300044998_151.tar.gz |
| LMW ETR NZ 2018 Waikite S 2 MG | 3300044998_152.tar.gz |
| LMW ETR NZ 2018 Waikite S 2 MG | 3300044998_157.tar.gz |
| LMW ETR NZ 2018 Waikite S 2 MG | 3300044998_159.tar.gz |
| LMW ETR NZ 2018 Waikite S 2 MG | 3300044998_16.tar.gz |
| LMW ETR NZ 2018 Waikite S 2 MG | 3300044998_160.tar.gz |
| LMW ETR NZ 2018 Waikite S 2 MG | 3300044998_163.tar.gz |
| LMW ETR NZ 2018 Waikite S 2 MG | 3300044998_164.tar.gz |
| LMW ETR NZ 2018 Waikite S 2 MG | 3300044998_166.tar.gz |
| LMW ETR NZ 2018 Waikite S 2 MG | 3300044998_169.tar.gz |
| LMW ETR NZ 2018 Waikite S 2 MG | 3300044998_170.tar.gz |
| LMW ETR NZ 2018 Waikite S 2 MG | 3300044998_171.tar.gz |
| LMW ETR NZ 2018 Waikite S 2 MG | 3300044998_172.tar.gz |
| LMW ETR NZ 2018 Waikite S 2 MG | 3300044998_174.tar.gz |
| LMW ETR NZ 2018 Waikite S 2 MG | 3300044998_176.tar.gz |
| LMW ETR NZ 2018 Waikite S 2 MG | 3300044998_177.tar.gz |
| LMW ETR NZ 2018 Waikite S 2 MG | 3300044998_179.tar.gz |
| LMW ETR NZ 2018 Waikite S 2 MG | 3300044998_18.tar.gz |
| LMW ETR NZ 2018 Waikite S 2 MG | 3300044998_180.tar.gz |
| LMW ETR NZ 2018 Waikite S 2 MG | 3300044998_183.tar.gz |
| LMW ETR NZ 2018 Waikite S 2 MG | 3300044998_184.tar.gz |
| LMW ETR NZ 2018 Waikite S 2 MG | 3300044998_185.tar.gz |
| LMW ETR NZ 2018 Waikite S 2 MG | 3300044998_187.tar.gz |
| LMW ETR NZ 2018 Waikite S 2 MG | 3300044998_188.tar.gz |
| LMW ETR NZ 2018 Waikite S 2 MG | 3300044998_189.tar.gz |
| LMW ETR NZ 2018 Waikite S 2 MG | 3300044998_19.tar.gz |
| LMW ETR NZ 2018 Waikite S 2 MG | 3300044998_190.tar.gz |
| LMW ETR NZ 2018 Waikite S 2 MG | 3300044998_191.tar.gz |
| LMW ETR NZ 2018 Waikite S 2 MG | 3300044998_192.tar.gz |
| LMW ETR NZ 2018 Waikite S 2 MG | 3300044998_194.tar.gz |
| LMW ETR NZ 2018 Waikite S 2 MG | 3300044998_196.tar.gz |
| LMW ETR NZ 2018 Waikite S 2 MG | 3300044998_197.tar.gz |
| LMW ETR NZ 2018 Waikite S 2 MG | 3300044998_198.tar.gz |
| LMW ETR NZ 2018 Waikite S 2 MG | 3300044998_199.tar.gz |
| LMW ETR NZ 2018 Waikite S 2 MG | 3300044998_2.tar.gz |
| LMW ETR NZ 2018 Waikite S 2 MG | 3300044998_20.tar.gz |
| LMW ETR NZ 2018 Waikite S 2 MG | 3300044998_200.tar.gz |
| LMW ETR NZ 2018 Waikite S 2 MG | 3300044998_201.tar.gz |
| LMW ETR NZ 2018 Waikite S 2 MG | 3300044998_202.tar.gz |
| LMW ETR NZ 2018 Waikite S 2 MG | 3300044998_203.tar.gz |
| LMW ETR NZ 2018 Waikite S 2 MG | 3300044998_204.tar.gz |
| LMW ETR NZ 2018 Waikite S 2 MG | 3300044998_205.tar.gz |
| LMW ETR NZ 2018 Waikite S 2 MG | 3300044998_206.tar.gz |
| LMW ETR NZ 2018 Waikite S 2 MG | 3300044998_208.tar.gz |
| LMW ETR NZ 2018 Waikite S 2 MG | 3300044998_21.tar.gz |
| LMW ETR NZ 2018 Waikite S 2 MG | 3300044998_213.tar.gz |
| LMW ETR NZ 2018 Waikite S 2 MG | 3300044998_214.tar.gz |
| LMW ETR NZ 2018 Waikite S 2 MG | 3300044998_215.tar.gz |
| LMW ETR NZ 2018 Waikite S 2 MG | 3300044998_216.tar.gz |
| LMW ETR NZ 2018 Waikite S 2 MG | 3300044998_217.tar.gz |
| LMW ETR NZ 2018 Waikite S 2 MG | 3300044998_23.tar.gz |
| LMW ETR NZ 2018 Waikite S 2 MG | 3300044998_24.tar.gz |
| LMW ETR NZ 2018 Waikite S 2 MG | 3300044998_25.tar.gz |
| LMW ETR NZ 2018 Waikite S 2 MG | 3300044998_26.tar.gz |
| LMW ETR NZ 2018 Waikite S 2 MG | 3300044998_27.tar.gz |
| LMW ETR NZ 2018 Waikite S 2 MG | 3300044998_3.tar.gz |
| LMW ETR NZ 2018 Waikite S 2 MG | 3300044998_30.tar.gz |
| LMW ETR NZ 2018 Waikite S 2 MG | 3300044998_32.tar.gz |
| LMW ETR NZ 2018 Waikite S 2 MG | 3300044998_4.tar.gz |
| LMW ETR NZ 2018 Waikite S 2 MG | 3300044998_40.tar.gz |
| LMW ETR NZ 2018 Waikite S 2 MG | 3300044998_42.tar.gz |
| LMW ETR NZ 2018 Waikite S 2 MG | 3300044998_45.tar.gz |
| LMW ETR NZ 2018 Waikite S 2 MG | 3300044998_49.tar.gz |
| LMW ETR NZ 2018 Waikite S 2 MG | 3300044998_5.tar.gz |
| LMW ETR NZ 2018 Waikite S 2 MG | 3300044998_53.tar.gz |
| LMW ETR NZ 2018 Waikite S 2 MG | 3300044998_58.tar.gz |
| LMW ETR NZ 2018 Waikite S 2 MG | 3300044998_6.tar.gz |
| LMW ETR NZ 2018 Waikite S 2 MG | 3300044998_61.tar.gz |
| LMW ETR NZ 2018 Waikite S 2 MG | 3300044998_62.tar.gz |
| LMW ETR NZ 2018 Waikite S 2 MG | 3300044998_63.tar.gz |
| LMW ETR NZ 2018 Waikite S 2 MG | 3300044998_65.tar.gz |
| LMW ETR NZ 2018 Waikite S 2 MG | 3300044998_67.tar.gz |
| LMW ETR NZ 2018 Waikite S 2 MG | 3300044998_68.tar.gz |
| LMW ETR NZ 2018 Waikite S 2 MG | 3300044998_69.tar.gz |
| LMW ETR NZ 2018 Waikite S 2 MG | 3300044998_70.tar.gz |
| LMW ETR NZ 2018 Waikite S 2 MG | 3300044998_72.tar.gz |
| LMW ETR NZ 2018 Waikite S 2 MG | 3300044998_76.tar.gz |
| LMW ETR NZ 2018 Waikite S 2 MG | 3300044998_79.tar.gz |
| LMW ETR NZ 2018 Waikite S 2 MG | 3300044998_8.tar.gz |
| LMW ETR NZ 2018 Waikite S 2 MG | 3300044998_80.tar.gz |
| LMW ETR NZ 2018 Waikite S 2 MG | 3300044998_82.tar.gz |
| LMW ETR NZ 2018 Waikite S 2 MG | 3300044998_84.tar.gz |
| LMW ETR NZ 2018 Waikite S 2 MG | 3300044998_87.tar.gz |
| LMW ETR NZ 2018 Waikite S 2 MG | 3300044998_88.tar.gz |
| LMW ETR NZ 2018 Waikite S 2 MG | 3300044998_89.tar.gz |
| LMW ETR NZ 2018 Waikite S 2 MG | 3300044998_90.tar.gz |
| LMW ETR NZ 2018 Waikite S 2 MG | 3300044998_92.tar.gz |
| LMW ETR NZ 2018 Waikite S 2 MG | 3300044998_94.tar.gz |
| LMW ETR NZ 2018 Waikite S 2 MG | 3300044998_95.tar.gz |
| LMW ETR NZ 2018 Waikite S 2 MG | 3300044998_96.tar.gz |
| LMW ETR NZ 2018 Waikite S 2 MG | 3300044998_98.tar.gz |

| Sequencing Project Name | File Name |
|---|---|
| LMW ETR NZ 2018 Waikite 7 2 MG | 3300045902_100.tar.gz |
| LMW ETR NZ 2018 Waikite 7 2 MG | 3300045902_101.tar.gz |
| LMW ETR NZ 2018 Waikite 7 2 MG | 3300045902_102.tar.gz |
| LMW ETR NZ 2018 Waikite 7 2 MG | 3300045902_105.tar.gz |
| LMW ETR NZ 2018 Waikite 7 2 MG | 3300045902_107.tar.gz |
| LMW ETR NZ 2018 Waikite 7 2 MG | 3300045902_108.tar.gz |
| LMW ETR NZ 2018 Waikite 7 2 MG | 3300045902_109.tar.gz |
| LMW ETR NZ 2018 Waikite 7 2 MG | 3300045902_11.tar.gz |
| LMW ETR NZ 2018 Waikite 7 2 MG | 3300045902_111.tar.gz |
| LMW ETR NZ 2018 Waikite 7 2 MG | 3300045902_115.tar.gz |
| LMW ETR NZ 2018 Waikite 7 2 MG | 3300045902_117.tar.gz |
| LMW ETR NZ 2018 Waikite 7 2 MG | 3300045902_119.tar.gz |
| LMW ETR NZ 2018 Waikite 7 2 MG | 3300045902_121.tar.gz |
| LMW ETR NZ 2018 Waikite 7 2 MG | 3300045902_124.tar.gz |
| LMW ETR NZ 2018 Waikite 7 2 MG | 3300045902_125.tar.gz |
| LMW ETR NZ 2018 Waikite 7 2 MG | 3300045902_126.tar.gz |
| LMW ETR NZ 2018 Waikite 7 2 MG | 3300045902_127.tar.gz |
| LMW ETR NZ 2018 Waikite 7 2 MG | 3300045902_128.tar.gz |
| LMW ETR NZ 2018 Waikite 7 2 MG | 3300045902_132.tar.gz |
| LMW ETR NZ 2018 Waikite 7 2 MG | 3300045902_133.tar.gz |
| LMW ETR NZ 2018 Waikite 7 2 MG | 3300045902_134.tar.gz |
| LMW ETR NZ 2018 Waikite 7 2 MG | 3300045902_136.tar.gz |
| LMW ETR NZ 2018 Waikite 7 2 MG | 3300045902_137.tar.gz |
| LMW ETR NZ 2018 Waikite 7 2 MG | 3300045902_139.tar.gz |
| LMW ETR NZ 2018 Waikite 7 2 MG | 3300045902_14.tar.gz |
| LMW ETR NZ 2018 Waikite 7 2 MG | 3300045902_140.tar.gz |
| LMW ETR NZ 2018 Waikite 7 2 MG | 3300045902_142.tar.gz |
| LMW ETR NZ 2018 Waikite 7 2 MG | 3300045902_143.tar.gz |
| LMW ETR NZ 2018 Waikite 7 2 MG | 3300045902_144.tar.gz |
| LMW ETR NZ 2018 Waikite 7 2 MG | 3300045902_145.tar.gz |
| LMW ETR NZ 2018 Waikite 7 2 MG | 3300045902_146.tar.gz |
| LMW ETR NZ 2018 Waikite 7 2 MG | 3300045902_149.tar.gz |
| LMW ETR NZ 2018 Waikite 7 2 MG | 3300045902_15.tar.gz |
| LMW ETR NZ 2018 Waikite 7 2 MG | 3300045902_150.tar.gz |
| LMW ETR NZ 2018 Waikite 7 2 MG | 3300045902_151.tar.gz |
| LMW ETR NZ 2018 Waikite 7 2 MG | 3300045902_153.tar.gz |
| LMW ETR NZ 2018 Waikite 7 2 MG | 3300045902_17.tar.gz |
| LMW ETR NZ 2018 Waikite 7 2 MG | 3300045902_18.tar.gz |
| LMW ETR NZ 2018 Waikite 7 2 MG | 3300045902_19.tar.gz |
| LMW ETR NZ 2018 Waikite 7 2 MG | 3300045902_2.tar.gz |
| LMW ETR NZ 2018 Waikite 7 2 MG | 3300045902_26.tar.gz |
| LMW ETR NZ 2018 Waikite 7 2 MG | 3300045902_29.tar.gz |
| LMW ETR NZ 2018 Waikite 7 2 MG | 3300045902_3.tar.gz |
| LMW ETR NZ 2018 Waikite 7 2 MG | 3300045902_30.tar.gz |
| LMW ETR NZ 2018 Waikite 7 2 MG | 3300045902_32.tar.gz |
| LMW ETR NZ 2018 Waikite 7 2 MG | 3300045902_33.tar.gz |
| LMW ETR NZ 2018 Waikite 7 2 MG | 3300045902_36.tar.gz |
| LMW ETR NZ 2018 Waikite 7 2 MG | 3300045902_38.tar.gz |
| LMW ETR NZ 2018 Waikite 7 2 MG | 3300045902_39.tar.gz |
| LMW ETR NZ 2018 Waikite 7 2 MG | 3300045902_4.tar.gz |
| LMW ETR NZ 2018 Waikite 7 2 MG | 3300045902_41.tar.gz |
| LMW ETR NZ 2018 Waikite 7 2 MG | 3300045902_42.tar.gz |
| LMW ETR NZ 2018 Waikite 7 2 MG | 3300045902_43.tar.gz |
| LMW ETR NZ 2018 Waikite 7 2 MG | 3300045902_44.tar.gz |
| LMW ETR NZ 2018 Waikite 7 2 MG | 3300045902_46.tar.gz |
| LMW ETR NZ 2018 Waikite 7 2 MG | 3300045902_48.tar.gz |
| LMW ETR NZ 2018 Waikite 7 2 MG | 3300045902_49.tar.gz |
| LMW ETR NZ 2018 Waikite 7 2 MG | 3300045902_50.tar.gz |
| LMW ETR NZ 2018 Waikite 7 2 MG | 3300045902_51.tar.gz |
| LMW ETR NZ 2018 Waikite 7 2 MG | 3300045902_52.tar.gz |
| LMW ETR NZ 2018 Waikite 7 2 MG | 3300045902_54.tar.gz |
| LMW ETR NZ 2018 Waikite 7 2 MG | 3300045902_55.tar.gz |
| LMW ETR NZ 2018 Waikite 7 2 MG | 3300045902_56.tar.gz |
| LMW ETR NZ 2018 Waikite 7 2 MG | 3300045902_58.tar.gz |
| LMW ETR NZ 2018 Waikite 7 2 MG | 3300045902_6.tar.gz |
| LMW ETR NZ 2018 Waikite 7 2 MG | 3300045902_66.tar.gz |
| LMW ETR NZ 2018 Waikite 7 2 MG | 3300045902_67.tar.gz |
| LMW ETR NZ 2018 Waikite 7 2 MG | 3300045902_68.tar.gz |
| LMW ETR NZ 2018 Waikite 7 2 MG | 3300045902_7.tar.gz |
| LMW ETR NZ 2018 Waikite 7 2 MG | 3300045902_74.tar.gz |
| LMW ETR NZ 2018 Waikite 7 2 MG | 3300045902_75.tar.gz |
| LMW ETR NZ 2018 Waikite 7 2 MG | 3300045902_78.tar.gz |
| LMW ETR NZ 2018 Waikite 7 2 MG | 3300045902_79.tar.gz |
| LMW ETR NZ 2018 Waikite 7 2 MG | 3300045902_8.tar.gz |
| LMW ETR NZ 2018 Waikite 7 2 MG | 3300045902_80.tar.gz |
| LMW ETR NZ 2018 Waikite 7 2 MG | 3300045902_81.tar.gz |
| LMW ETR NZ 2018 Waikite 7 2 MG | 3300045902_82.tar.gz |
| LMW ETR NZ 2018 Waikite 7 2 MG | 3300045902_83.tar.gz |
| LMW ETR NZ 2018 Waikite 7 2 MG | 3300045902_86.tar.gz |
| LMW ETR NZ 2018 Waikite 7 2 MG | 3300045902_87.tar.gz |
| LMW ETR NZ 2018 Waikite 7 2 MG | 3300045902_89.tar.gz |
| LMW ETR NZ 2018 Waikite 7 2 MG | 3300045902_90.tar.gz |
| LMW ETR NZ 2018 Waikite 7 2 MG | 3300045902_91.tar.gz |
| LMW ETR NZ 2018 Waikite 7 2 MG | 3300045902_93.tar.gz |
| LMW ETR NZ 2018 Waikite 7 2 MG | 3300045902_94.tar.gz |
| LMW ETR NZ 2018 Waikite 7 2 MG | 3300045902_96.tar.gz |
| LMW ETR NZ 2018 Waikite 7 2 MG | 3300045902_97.tar.gz |
| LMW ETR NZ 2018 Waikite 7 2 MG | 3300045902_98.tar.gz |

| Sequencing Project Name | File Name |
| --- | --- |
| LMW ETR NZ 2018 Waikite 7 1 MG | 3300045912_1.tar.gz |
| LMW ETR NZ 2018 Waikite 7 1 MG | 3300045912_103.tar.gz |
| LMW ETR NZ 2018 Waikite 7 1 MG | 3300045912_104.tar.gz |
| LMW ETR NZ 2018 Waikite 7 1 MG | 3300045912_105.tar.gz |
| LMW ETR NZ 2018 Waikite 7 1 MG | 3300045912_106.tar.gz |
| LMW ETR NZ 2018 Waikite 7 1 MG | 3300045912_107.tar.gz |
| LMW ETR NZ 2018 Waikite 7 1 MG | 3300045912_108.tar.gz |
| LMW ETR NZ 2018 Waikite 7 1 MG | 3300045912_109.tar.gz |
| LMW ETR NZ 2018 Waikite 7 1 MG | 3300045912_11.tar.gz |
| LMW ETR NZ 2018 Waikite 7 1 MG | 3300045912_111.tar.gz |
| LMW ETR NZ 2018 Waikite 7 1 MG | 3300045912_112.tar.gz |
| LMW ETR NZ 2018 Waikite 7 1 MG | 3300045912_113.tar.gz |
| LMW ETR NZ 2018 Waikite 7 1 MG | 3300045912_115.tar.gz |
| LMW ETR NZ 2018 Waikite 7 1 MG | 3300045912_118.tar.gz |
| LMW ETR NZ 2018 Waikite 7 1 MG | 3300045912_119.tar.gz |
| LMW ETR NZ 2018 Waikite 7 1 MG | 3300045912_120.tar.gz |
| LMW ETR NZ 2018 Waikite 7 1 MG | 3300045912_121.tar.gz |
| LMW ETR NZ 2018 Waikite 7 1 MG | 3300045912_125.tar.gz |
| LMW ETR NZ 2018 Waikite 7 1 MG | 3300045912_130.tar.gz |
| LMW ETR NZ 2018 Waikite 7 1 MG | 3300045912_131.tar.gz |
| LMW ETR NZ 2018 Waikite 7 1 MG | 3300045912_132.tar.gz |
| LMW ETR NZ 2018 Waikite 7 1 MG | 3300045912_133.tar.gz |
| LMW ETR NZ 2018 Waikite 7 1 MG | 3300045912_134.tar.gz |
| LMW ETR NZ 2018 Waikite 7 1 MG | 3300045912_137.tar.gz |
| LMW ETR NZ 2018 Waikite 7 1 MG | 3300045912_139.tar.gz |
| LMW ETR NZ 2018 Waikite 7 1 MG | 3300045912_14.tar.gz |
| LMW ETR NZ 2018 Waikite 7 1 MG | 3300045912_141.tar.gz |
| LMW ETR NZ 2018 Waikite 7 1 MG | 3300045912_142.tar.gz |
| LMW ETR NZ 2018 Waikite 7 1 MG | 3300045912_144.tar.gz |
| LMW ETR NZ 2018 Waikite 7 1 MG | 3300045912_146.tar.gz |
| LMW ETR NZ 2018 Waikite 7 1 MG | 3300045912_147.tar.gz |
| LMW ETR NZ 2018 Waikite 7 1 MG | 3300045912_15.tar.gz |
| LMW ETR NZ 2018 Waikite 7 1 MG | 3300045912_150.tar.gz |
| LMW ETR NZ 2018 Waikite 7 1 MG | 3300045912_151.tar.gz |
| LMW ETR NZ 2018 Waikite 7 1 MG | 3300045912_153.tar.gz |
| LMW ETR NZ 2018 Waikite 7 1 MG | 3300045912_156.tar.gz |
| LMW ETR NZ 2018 Waikite 7 1 MG | 3300045912_157.tar.gz |
| LMW ETR NZ 2018 Waikite 7 1 MG | 3300045912_158.tar.gz |
| LMW ETR NZ 2018 Waikite 7 1 MG | 3300045912_159.tar.gz |
| LMW ETR NZ 2018 Waikite 7 1 MG | 3300045912_160.tar.gz |
| LMW ETR NZ 2018 Waikite 7 1 MG | 3300045912_163.tar.gz |
| LMW ETR NZ 2018 Waikite 7 1 MG | 3300045912_166.tar.gz |
| LMW ETR NZ 2018 Waikite 7 1 MG | 3300045912_167.tar.gz |
| LMW ETR NZ 2018 Waikite 7 1 MG | 3300045912_168.tar.gz |
| LMW ETR NZ 2018 Waikite 7 1 MG | 3300045912_17.tar.gz |
| LMW ETR NZ 2018 Waikite 7 1 MG | 3300045912_170.tar.gz |
| LMW ETR NZ 2018 Waikite 7 1 MG | 3300045912_171.tar.gz |
| LMW ETR NZ 2018 Waikite 7 1 MG | 3300045912_172.tar.gz |
| LMW ETR NZ 2018 Waikite 7 1 MG | 3300045912_173.tar.gz |
| LMW ETR NZ 2018 Waikite 7 1 MG | 3300045912_174.tar.gz |
| LMW ETR NZ 2018 Waikite 7 1 MG | 3300045912_176.tar.gz |
| LMW ETR NZ 2018 Waikite 7 1 MG | 3300045912_177.tar.gz |
| LMW ETR NZ 2018 Waikite 7 1 MG | 3300045912_18.tar.gz |
| LMW ETR NZ 2018 Waikite 7 1 MG | 3300045912_2.tar.gz |
| LMW ETR NZ 2018 Waikite 7 1 MG | 3300045912_20.tar.gz |
| LMW ETR NZ 2018 Waikite 7 1 MG | 3300045912_21.tar.gz |
| LMW ETR NZ 2018 Waikite 7 1 MG | 3300045912_23.tar.gz |
| LMW ETR NZ 2018 Waikite 7 1 MG | 3300045912_25.tar.gz |
| LMW ETR NZ 2018 Waikite 7 1 MG | 3300045912_26.tar.gz |
| LMW ETR NZ 2018 Waikite 7 1 MG | 3300045912_29.tar.gz |
| LMW ETR NZ 2018 Waikite 7 1 MG | 3300045912_3.tar.gz |
| LMW ETR NZ 2018 Waikite 7 1 MG | 3300045912_32.tar.gz |
| LMW ETR NZ 2018 Waikite 7 1 MG | 3300045912_36.tar.gz |
| LMW ETR NZ 2018 Waikite 7 1 MG | 3300045912_37.tar.gz |
| LMW ETR NZ 2018 Waikite 7 1 MG | 3300045912_38.tar.gz |
| LMW ETR NZ 2018 Waikite 7 1 MG | 3300045912_39.tar.gz |
| LMW ETR NZ 2018 Waikite 7 1 MG | 3300045912_40.tar.gz |
| LMW ETR NZ 2018 Waikite 7 1 MG | 3300045912_44.tar.gz |
| LMW ETR NZ 2018 Waikite 7 1 MG | 3300045912_46.tar.gz |
| LMW ETR NZ 2018 Waikite 7 1 MG | 3300045912_50.tar.gz |
| LMW ETR NZ 2018 Waikite 7 1 MG | 3300045912_53.tar.gz |
| LMW ETR NZ 2018 Waikite 7 1 MG | 3300045912_55.tar.gz |
| LMW ETR NZ 2018 Waikite 7 1 MG | 3300045912_56.tar.gz |
| LMW ETR NZ 2018 Waikite 7 1 MG | 3300045912_59.tar.gz |
| LMW ETR NZ 2018 Waikite 7 1 MG | 3300045912_6.tar.gz |
| LMW ETR NZ 2018 Waikite 7 1 MG | 3300045912_60.tar.gz |
| LMW ETR NZ 2018 Waikite 7 1 MG | 3300045912_62.tar.gz |
| LMW ETR NZ 2018 Waikite 7 1 MG | 3300045912_64.tar.gz |
| LMW ETR NZ 2018 Waikite 7 1 MG | 3300045912_65.tar.gz |
| LMW ETR NZ 2018 Waikite 7 1 MG | 3300045912_66.tar.gz |
| LMW ETR NZ 2018 Waikite 7 1 MG | 3300045912_67.tar.gz |
| LMW ETR NZ 2018 Waikite 7 1 MG | 3300045912_69.tar.gz |
| LMW ETR NZ 2018 Waikite 7 1 MG | 3300045912_71.tar.gz |
| LMW ETR NZ 2018 Waikite 7 1 MG | 3300045912_73.tar.gz |
| LMW ETR NZ 2018 Waikite 7 1 MG | 3300045912_74.tar.gz |
| LMW ETR NZ 2018 Waikite 7 1 MG | 3300045912_75.tar.gz |
| LMW ETR NZ 2018 Waikite 7 1 MG | 3300045912_77.tar.gz |
| LMW ETR NZ 2018 Waikite 7 1 MG | 3300045912_79.tar.gz |
| LMW ETR NZ 2018 Waikite 7 1 MG | 3300045912_8.tar.gz |
| LMW ETR NZ 2018 Waikite 7 1 MG | 3300045912_80.tar.gz |
| LMW ETR NZ 2018 Waikite 7 1 MG | 3300045912_82.tar.gz |
| LMW ETR NZ 2018 Waikite 7 1 MG | 3300045912_83.tar.gz |
| LMW ETR NZ 2018 Waikite 7 1 MG | 3300045912_85.tar.gz |
| LMW ETR NZ 2018 Waikite 7 1 MG | 3300045912_86.tar.gz |
| LMW ETR NZ 2018 Waikite 7 1 MG | 3300045912_87.tar.gz |
| LMW ETR NZ 2018 Waikite 7 1 MG | 3300045912_88.tar.gz |
| LMW ETR NZ 2018 Waikite 7 1 MG | 3300045912_89.tar.gz |
| LMW ETR NZ 2018 Waikite 7 1 MG | 3300045912_90.tar.gz |
| LMW ETR NZ 2018 Waikite 7 1 MG | 3300045912_91.tar.gz |
| LMW ETR NZ 2018 Waikite 7 1 MG | 3300045912_92.tar.gz |
| LMW ETR NZ 2018 Waikite 7 1 MG | 3300045912_93.tar.gz |
| LMW ETR NZ 2018 Waikite 7 1 MG | 3300045912_94.tar.gz |
| LMW ETR NZ 2018 Waikite 7 1 MG | 3300045912_95.tar.gz |
| LMW ETR NZ 2018 Waikite 7 1 MG | 3300045912_96.tar.gz |

| LMW ETR NZ 2018 Waikite 7 1 MG | 3300045912_97.tar.gz |
| LMW ETR NZ 2018 Waikite 7 1 MG | 3300045912_98.tar.gz |
| LMW ETR NZ 2018 Waikite 7 1 MG | 3300045912_99.tar.gz |

| Sequencing Project Name | File Name |
| --- | --- |
| LMW ETR NZ 2018 Waikite 6 2b MG | 3300044960_1.tar.gz |
| LMW ETR NZ 2018 Waikite 6 2b MG | 3300044960_101.tar.gz |
| LMW ETR NZ 2018 Waikite 6 2b MG | 3300044960_102.tar.gz |
| LMW ETR NZ 2018 Waikite 6 2b MG | 3300044960_103.tar.gz |
| LMW ETR NZ 2018 Waikite 6 2b MG | 3300044960_105.tar.gz |
| LMW ETR NZ 2018 Waikite 6 2b MG | 3300044960_106.tar.gz |
| LMW ETR NZ 2018 Waikite 6 2b MG | 3300044960_107.tar.gz |
| LMW ETR NZ 2018 Waikite 6 2b MG | 3300044960_108.tar.gz |
| LMW ETR NZ 2018 Waikite 6 2b MG | 3300044960_109.tar.gz |
| LMW ETR NZ 2018 Waikite 6 2b MG | 3300044960_11.tar.gz |
| LMW ETR NZ 2018 Waikite 6 2b MG | 3300044960_110.tar.gz |
| LMW ETR NZ 2018 Waikite 6 2b MG | 3300044960_111.tar.gz |
| LMW ETR NZ 2018 Waikite 6 2b MG | 3300044960_112.tar.gz |
| LMW ETR NZ 2018 Waikite 6 2b MG | 3300044960_113.tar.gz |
| LMW ETR NZ 2018 Waikite 6 2b MG | 3300044960_115.tar.gz |
| LMW ETR NZ 2018 Waikite 6 2b MG | 3300044960_116.tar.gz |
| LMW ETR NZ 2018 Waikite 6 2b MG | 3300044960_117.tar.gz |
| LMW ETR NZ 2018 Waikite 6 2b MG | 3300044960_118.tar.gz |
| LMW ETR NZ 2018 Waikite 6 2b MG | 3300044960_12.tar.gz |
| LMW ETR NZ 2018 Waikite 6 2b MG | 3300044960_120.tar.gz |
| LMW ETR NZ 2018 Waikite 6 2b MG | 3300044960_121.tar.gz |
| LMW ETR NZ 2018 Waikite 6 2b MG | 3300044960_122.tar.gz |
| LMW ETR NZ 2018 Waikite 6 2b MG | 3300044960_123.tar.gz |
| LMW ETR NZ 2018 Waikite 6 2b MG | 3300044960_124.tar.gz |
| LMW ETR NZ 2018 Waikite 6 2b MG | 3300044960_127.tar.gz |
| LMW ETR NZ 2018 Waikite 6 2b MG | 3300044960_128.tar.gz |
| LMW ETR NZ 2018 Waikite 6 2b MG | 3300044960_129.tar.gz |
| LMW ETR NZ 2018 Waikite 6 2b MG | 3300044960_130.tar.gz |
| LMW ETR NZ 2018 Waikite 6 2b MG | 3300044960_135.tar.gz |
| LMW ETR NZ 2018 Waikite 6 2b MG | 3300044960_136.tar.gz |
| LMW ETR NZ 2018 Waikite 6 2b MG | 3300044960_14.tar.gz |
| LMW ETR NZ 2018 Waikite 6 2b MG | 3300044960_140.tar.gz |
| LMW ETR NZ 2018 Waikite 6 2b MG | 3300044960_144.tar.gz |
| LMW ETR NZ 2018 Waikite 6 2b MG | 3300044960_145.tar.gz |
| LMW ETR NZ 2018 Waikite 6 2b MG | 3300044960_147.tar.gz |
| LMW ETR NZ 2018 Waikite 6 2b MG | 3300044960_148.tar.gz |
| LMW ETR NZ 2018 Waikite 6 2b MG | 3300044960_15.tar.gz |
| LMW ETR NZ 2018 Waikite 6 2b MG | 3300044960_153.tar.gz |
| LMW ETR NZ 2018 Waikite 6 2b MG | 3300044960_154.tar.gz |
| LMW ETR NZ 2018 Waikite 6 2b MG | 3300044960_155.tar.gz |
| LMW ETR NZ 2018 Waikite 6 2b MG | 3300044960_156.tar.gz |
| LMW ETR NZ 2018 Waikite 6 2b MG | 3300044960_157.tar.gz |
| LMW ETR NZ 2018 Waikite 6 2b MG | 3300044960_16.tar.gz |
| LMW ETR NZ 2018 Waikite 6 2b MG | 3300044960_161.tar.gz |
| LMW ETR NZ 2018 Waikite 6 2b MG | 3300044960_163.tar.gz |
| LMW ETR NZ 2018 Waikite 6 2b MG | 3300044960_165.tar.gz |
| LMW ETR NZ 2018 Waikite 6 2b MG | 3300044960_169.tar.gz |
| LMW ETR NZ 2018 Waikite 6 2b MG | 3300044960_170.tar.gz |
| LMW ETR NZ 2018 Waikite 6 2b MG | 3300044960_171.tar.gz |
| LMW ETR NZ 2018 Waikite 6 2b MG | 3300044960_173.tar.gz |
| LMW ETR NZ 2018 Waikite 6 2b MG | 3300044960_175.tar.gz |
| LMW ETR NZ 2018 Waikite 6 2b MG | 3300044960_177.tar.gz |
| LMW ETR NZ 2018 Waikite 6 2b MG | 3300044960_178.tar.gz |
| LMW ETR NZ 2018 Waikite 6 2b MG | 3300044960_181.tar.gz |
| LMW ETR NZ 2018 Waikite 6 2b MG | 3300044960_182.tar.gz |
| LMW ETR NZ 2018 Waikite 6 2b MG | 3300044960_183.tar.gz |
| LMW ETR NZ 2018 Waikite 6 2b MG | 3300044960_184.tar.gz |
| LMW ETR NZ 2018 Waikite 6 2b MG | 3300044960_185.tar.gz |
| LMW ETR NZ 2018 Waikite 6 2b MG | 3300044960_186.tar.gz |
| LMW ETR NZ 2018 Waikite 6 2b MG | 3300044960_19.tar.gz |
| LMW ETR NZ 2018 Waikite 6 2b MG | 3300044960_190.tar.gz |
| LMW ETR NZ 2018 Waikite 6 2b MG | 3300044960_2.tar.gz |
| LMW ETR NZ 2018 Waikite 6 2b MG | 3300044960_21.tar.gz |
| LMW ETR NZ 2018 Waikite 6 2b MG | 3300044960_22.tar.gz |
| LMW ETR NZ 2018 Waikite 6 2b MG | 3300044960_23.tar.gz |
| LMW ETR NZ 2018 Waikite 6 2b MG | 3300044960_24.tar.gz |
| LMW ETR NZ 2018 Waikite 6 2b MG | 3300044960_27.tar.gz |
| LMW ETR NZ 2018 Waikite 6 2b MG | 3300044960_28.tar.gz |
| LMW ETR NZ 2018 Waikite 6 2b MG | 3300044960_29.tar.gz |
| LMW ETR NZ 2018 Waikite 6 2b MG | 3300044960_33.tar.gz |
| LMW ETR NZ 2018 Waikite 6 2b MG | 3300044960_36.tar.gz |
| LMW ETR NZ 2018 Waikite 6 2b MG | 3300044960_37.tar.gz |
| LMW ETR NZ 2018 Waikite 6 2b MG | 3300044960_38.tar.gz |
| LMW ETR NZ 2018 Waikite 6 2b MG | 3300044960_39.tar.gz |
| LMW ETR NZ 2018 Waikite 6 2b MG | 3300044960_4.tar.gz |
| LMW ETR NZ 2018 Waikite 6 2b MG | 3300044960_40.tar.gz |
| LMW ETR NZ 2018 Waikite 6 2b MG | 3300044960_43.tar.gz |
| LMW ETR NZ 2018 Waikite 6 2b MG | 3300044960_44.tar.gz |
| LMW ETR NZ 2018 Waikite 6 2b MG | 3300044960_47.tar.gz |
| LMW ETR NZ 2018 Waikite 6 2b MG | 3300044960_5.tar.gz |
| LMW ETR NZ 2018 Waikite 6 2b MG | 3300044960_52.tar.gz |
| LMW ETR NZ 2018 Waikite 6 2b MG | 3300044960_54.tar.gz |
| LMW ETR NZ 2018 Waikite 6 2b MG | 3300044960_55.tar.gz |
| LMW ETR NZ 2018 Waikite 6 2b MG | 3300044960_56.tar.gz |
| LMW ETR NZ 2018 Waikite 6 2b MG | 3300044960_57.tar.gz |
| LMW ETR NZ 2018 Waikite 6 2b MG | 3300044960_58.tar.gz |
| LMW ETR NZ 2018 Waikite 6 2b MG | 3300044960_59.tar.gz |
| LMW ETR NZ 2018 Waikite 6 2b MG | 3300044960_60.tar.gz |
| LMW ETR NZ 2018 Waikite 6 2b MG | 3300044960_61.tar.gz |
| LMW ETR NZ 2018 Waikite 6 2b MG | 3300044960_62.tar.gz |
| LMW ETR NZ 2018 Waikite 6 2b MG | 3300044960_63.tar.gz |
| LMW ETR NZ 2018 Waikite 6 2b MG | 3300044960_64.tar.gz |
| LMW ETR NZ 2018 Waikite 6 2b MG | 3300044960_66.tar.gz |
| LMW ETR NZ 2018 Waikite 6 2b MG | 3300044960_67.tar.gz |
| LMW ETR NZ 2018 Waikite 6 2b MG | 3300044960_68.tar.gz |
| LMW ETR NZ 2018 Waikite 6 2b MG | 3300044960_7.tar.gz |
| LMW ETR NZ 2018 Waikite 6 2b MG | 3300044960_70.tar.gz |
| LMW ETR NZ 2018 Waikite 6 2b MG | 3300044960_71.tar.gz |

| LMW ETR NZ 2018 Waikite 6 2b MG | 3300044960_72.tar.gz |
| LMW ETR NZ 2018 Waikite 6 2b MG | 3300044960_73.tar.gz |
| LMW ETR NZ 2018 Waikite 6 2b MG | 3300044960_75.tar.gz |
| LMW ETR NZ 2018 Waikite 6 2b MG | 3300044960_76.tar.gz |
| LMW ETR NZ 2018 Waikite 6 2b MG | 3300044960_78.tar.gz |
| LMW ETR NZ 2018 Waikite 6 2b MG | 3300044960_79.tar.gz |
| LMW ETR NZ 2018 Waikite 6 2b MG | 3300044960_80.tar.gz |
| LMW ETR NZ 2018 Waikite 6 2b MG | 3300044960_81.tar.gz |
| LMW ETR NZ 2018 Waikite 6 2b MG | 3300044960_82.tar.gz |
| LMW ETR NZ 2018 Waikite 6 2b MG | 3300044960_83.tar.gz |
| LMW ETR NZ 2018 Waikite 6 2b MG | 3300044960_84.tar.gz |
| LMW ETR NZ 2018 Waikite 6 2b MG | 3300044960_85.tar.gz |
| LMW ETR NZ 2018 Waikite 6 2b MG | 3300044960_86.tar.gz |
| LMW ETR NZ 2018 Waikite 6 2b MG | 3300044960_88.tar.gz |
| LMW ETR NZ 2018 Waikite 6 2b MG | 3300044960_89.tar.gz |
| LMW ETR NZ 2018 Waikite 6 2b MG | 3300044960_9.tar.gz |
| LMW ETR NZ 2018 Waikite 6 2b MG | 3300044960_90.tar.gz |
| LMW ETR NZ 2018 Waikite 6 2b MG | 3300044960_92.tar.gz |
| LMW ETR NZ 2018 Waikite 6 2b MG | 3300044960_93.tar.gz |
| LMW ETR NZ 2018 Waikite 6 2b MG | 3300044960_94.tar.gz |
| LMW ETR NZ 2018 Waikite 6 2b MG | 3300044960_96.tar.gz |
| LMW ETR NZ 2018 Waikite 6 2b MG | 3300044960_97.tar.gz |
| LMW ETR NZ 2018 Waikite 6 2b MG | 3300044960_98.tar.gz |
| LMW ETR NZ 2018 Waikite 6 2b MG | 3300044960_99.tar.gz |

| Sequencing Project Name | File Name |
| --- | --- |
| LMW ETR NZ 2018 Waikite 6 2a MG | 3300045136_1.tar.gz |
| LMW ETR NZ 2018 Waikite 6 2a MG | 3300045136_100.tar.gz |
| LMW ETR NZ 2018 Waikite 6 2a MG | 3300045136_102.tar.gz |
| LMW ETR NZ 2018 Waikite 6 2a MG | 3300045136_103.tar.gz |
| LMW ETR NZ 2018 Waikite 6 2a MG | 3300045136_106.tar.gz |
| LMW ETR NZ 2018 Waikite 6 2a MG | 3300045136_107.tar.gz |
| LMW ETR NZ 2018 Waikite 6 2a MG | 3300045136_110.tar.gz |
| LMW ETR NZ 2018 Waikite 6 2a MG | 3300045136_112.tar.gz |
| LMW ETR NZ 2018 Waikite 6 2a MG | 3300045136_113.tar.gz |
| LMW ETR NZ 2018 Waikite 6 2a MG | 3300045136_114.tar.gz |
| LMW ETR NZ 2018 Waikite 6 2a MG | 3300045136_115.tar.gz |
| LMW ETR NZ 2018 Waikite 6 2a MG | 3300045136_116.tar.gz |
| LMW ETR NZ 2018 Waikite 6 2a MG | 3300045136_12.tar.gz |
| LMW ETR NZ 2018 Waikite 6 2a MG | 3300045136_120.tar.gz |
| LMW ETR NZ 2018 Waikite 6 2a MG | 3300045136_123.tar.gz |
| LMW ETR NZ 2018 Waikite 6 2a MG | 3300045136_124.tar.gz |
| LMW ETR NZ 2018 Waikite 6 2a MG | 3300045136_125.tar.gz |
| LMW ETR NZ 2018 Waikite 6 2a MG | 3300045136_127.tar.gz |
| LMW ETR NZ 2018 Waikite 6 2a MG | 3300045136_128.tar.gz |
| LMW ETR NZ 2018 Waikite 6 2a MG | 3300045136_129.tar.gz |
| LMW ETR NZ 2018 Waikite 6 2a MG | 3300045136_13.tar.gz |
| LMW ETR NZ 2018 Waikite 6 2a MG | 3300045136_130.tar.gz |
| LMW ETR NZ 2018 Waikite 6 2a MG | 3300045136_131.tar.gz |
| LMW ETR NZ 2018 Waikite 6 2a MG | 3300045136_132.tar.gz |
| LMW ETR NZ 2018 Waikite 6 2a MG | 3300045136_133.tar.gz |
| LMW ETR NZ 2018 Waikite 6 2a MG | 3300045136_135.tar.gz |
| LMW ETR NZ 2018 Waikite 6 2a MG | 3300045136_137.tar.gz |
| LMW ETR NZ 2018 Waikite 6 2a MG | 3300045136_138.tar.gz |
| LMW ETR NZ 2018 Waikite 6 2a MG | 3300045136_14.tar.gz |
| LMW ETR NZ 2018 Waikite 6 2a MG | 3300045136_140.tar.gz |
| LMW ETR NZ 2018 Waikite 6 2a MG | 3300045136_141.tar.gz |
| LMW ETR NZ 2018 Waikite 6 2a MG | 3300045136_144.tar.gz |
| LMW ETR NZ 2018 Waikite 6 2a MG | 3300045136_145.tar.gz |
| LMW ETR NZ 2018 Waikite 6 2a MG | 3300045136_146.tar.gz |
| LMW ETR NZ 2018 Waikite 6 2a MG | 3300045136_148.tar.gz |
| LMW ETR NZ 2018 Waikite 6 2a MG | 3300045136_15.tar.gz |
| LMW ETR NZ 2018 Waikite 6 2a MG | 3300045136_150.tar.gz |
| LMW ETR NZ 2018 Waikite 6 2a MG | 3300045136_152.tar.gz |
| LMW ETR NZ 2018 Waikite 6 2a MG | 3300045136_153.tar.gz |
| LMW ETR NZ 2018 Waikite 6 2a MG | 3300045136_155.tar.gz |
| LMW ETR NZ 2018 Waikite 6 2a MG | 3300045136_158.tar.gz |
| LMW ETR NZ 2018 Waikite 6 2a MG | 3300045136_16.tar.gz |
| LMW ETR NZ 2018 Waikite 6 2a MG | 3300045136_160.tar.gz |
| LMW ETR NZ 2018 Waikite 6 2a MG | 3300045136_164.tar.gz |
| LMW ETR NZ 2018 Waikite 6 2a MG | 3300045136_166.tar.gz |
| LMW ETR NZ 2018 Waikite 6 2a MG | 3300045136_167.tar.gz |
| LMW ETR NZ 2018 Waikite 6 2a MG | 3300045136_17.tar.gz |
| LMW ETR NZ 2018 Waikite 6 2a MG | 3300045136_170.tar.gz |
| LMW ETR NZ 2018 Waikite 6 2a MG | 3300045136_171.tar.gz |
| LMW ETR NZ 2018 Waikite 6 2a MG | 3300045136_172.tar.gz |
| LMW ETR NZ 2018 Waikite 6 2a MG | 3300045136_173.tar.gz |
| LMW ETR NZ 2018 Waikite 6 2a MG | 3300045136_175.tar.gz |
| LMW ETR NZ 2018 Waikite 6 2a MG | 3300045136_176.tar.gz |
| LMW ETR NZ 2018 Waikite 6 2a MG | 3300045136_177.tar.gz |
| LMW ETR NZ 2018 Waikite 6 2a MG | 3300045136_179.tar.gz |
| LMW ETR NZ 2018 Waikite 6 2a MG | 3300045136_18.tar.gz |
| LMW ETR NZ 2018 Waikite 6 2a MG | 3300045136_182.tar.gz |
| LMW ETR NZ 2018 Waikite 6 2a MG | 3300045136_183.tar.gz |
| LMW ETR NZ 2018 Waikite 6 2a MG | 3300045136_187.tar.gz |
| LMW ETR NZ 2018 Waikite 6 2a MG | 3300045136_19.tar.gz |
| LMW ETR NZ 2018 Waikite 6 2a MG | 3300045136_2.tar.gz |
| LMW ETR NZ 2018 Waikite 6 2a MG | 3300045136_20.tar.gz |
| LMW ETR NZ 2018 Waikite 6 2a MG | 3300045136_21.tar.gz |
| LMW ETR NZ 2018 Waikite 6 2a MG | 3300045136_25.tar.gz |
| LMW ETR NZ 2018 Waikite 6 2a MG | 3300045136_26.tar.gz |
| LMW ETR NZ 2018 Waikite 6 2a MG | 3300045136_27.tar.gz |
| LMW ETR NZ 2018 Waikite 6 2a MG | 3300045136_29.tar.gz |
| LMW ETR NZ 2018 Waikite 6 2a MG | 3300045136_3.tar.gz |
| LMW ETR NZ 2018 Waikite 6 2a MG | 3300045136_31.tar.gz |
| LMW ETR NZ 2018 Waikite 6 2a MG | 3300045136_32.tar.gz |
| LMW ETR NZ 2018 Waikite 6 2a MG | 3300045136_34.tar.gz |
| LMW ETR NZ 2018 Waikite 6 2a MG | 3300045136_35.tar.gz |
| LMW ETR NZ 2018 Waikite 6 2a MG | 3300045136_36.tar.gz |
| LMW ETR NZ 2018 Waikite 6 2a MG | 3300045136_37.tar.gz |
| LMW ETR NZ 2018 Waikite 6 2a MG | 3300045136_39.tar.gz |
| LMW ETR NZ 2018 Waikite 6 2a MG | 3300045136_4.tar.gz |
| LMW ETR NZ 2018 Waikite 6 2a MG | 3300045136_42.tar.gz |
| LMW ETR NZ 2018 Waikite 6 2a MG | 3300045136_44.tar.gz |
| LMW ETR NZ 2018 Waikite 6 2a MG | 3300045136_45.tar.gz |
| LMW ETR NZ 2018 Waikite 6 2a MG | 3300045136_46.tar.gz |
| LMW ETR NZ 2018 Waikite 6 2a MG | 3300045136_47.tar.gz |
| LMW ETR NZ 2018 Waikite 6 2a MG | 3300045136_48.tar.gz |
| LMW ETR NZ 2018 Waikite 6 2a MG | 3300045136_49.tar.gz |
| LMW ETR NZ 2018 Waikite 6 2a MG | 3300045136_51.tar.gz |
| LMW ETR NZ 2018 Waikite 6 2a MG | 3300045136_54.tar.gz |
| LMW ETR NZ 2018 Waikite 6 2a MG | 3300045136_55.tar.gz |
| LMW ETR NZ 2018 Waikite 6 2a MG | 3300045136_56.tar.gz |
| LMW ETR NZ 2018 Waikite 6 2a MG | 3300045136_57.tar.gz |
| LMW ETR NZ 2018 Waikite 6 2a MG | 3300045136_59.tar.gz |
| LMW ETR NZ 2018 Waikite 6 2a MG | 3300045136_60.tar.gz |
| LMW ETR NZ 2018 Waikite 6 2a MG | 3300045136_62.tar.gz |
| LMW ETR NZ 2018 Waikite 6 2a MG | 3300045136_63.tar.gz |
| LMW ETR NZ 2018 Waikite 6 2a MG | 3300045136_64.tar.gz |
| LMW ETR NZ 2018 Waikite 6 2a MG | 3300045136_65.tar.gz |
| LMW ETR NZ 2018 Waikite 6 2a MG | 3300045136_66.tar.gz |
| LMW ETR NZ 2018 Waikite 6 2a MG | 3300045136_7.tar.gz |
| LMW ETR NZ 2018 Waikite 6 2a MG | 3300045136_70.tar.gz |
| LMW ETR NZ 2018 Waikite 6 2a MG | 3300045136_72.tar.gz |
| LMW ETR NZ 2018 Waikite 6 2a MG | 3300045136_74.tar.gz |
| LMW ETR NZ 2018 Waikite 6 2a MG | 3300045136_77.tar.gz |
| LMW ETR NZ 2018 Waikite 6 2a MG | 3300045136_80.tar.gz |
| LMW ETR NZ 2018 Waikite 6 2a MG | 3300045136_81.tar.gz |
| LMW ETR NZ 2018 Waikite 6 2a MG | 3300045136_82.tar.gz |
| LMW ETR NZ 2018 Waikite 6 2a MG | 3300045136_83.tar.gz |
| LMW ETR NZ 2018 Waikite 6 2a MG | 3300045136_84.tar.gz |
| LMW ETR NZ 2018 Waikite 6 2a MG | 3300045136_85.tar.gz |
| LMW ETR NZ 2018 Waikite 6 2a MG | 3300045136_86.tar.gz |
| LMW ETR NZ 2018 Waikite 6 2a MG | 3300045136_87.tar.gz |
| LMW ETR NZ 2018 Waikite 6 2a MG | 3300045136_88.tar.gz |
| LMW ETR NZ 2018 Waikite 6 2a MG | 3300045136_89.tar.gz |
| LMW ETR NZ 2018 Waikite 6 2a MG | 3300045136_91.tar.gz |
| LMW ETR NZ 2018 Waikite 6 2a MG | 3300045136_92.tar.gz |
| LMW ETR NZ 2018 Waikite 6 2a MG | 3300045136_96.tar.gz |
| LMW ETR NZ 2018 Waikite 6 2a MG | 3300045136_97.tar.gz |
| LMW ETR NZ 2018 Waikite 6 2a MG | 3300045136_98.tar.gz |
| LMW ETR NZ 2018 Waikite 6 2a MG | 3300045136_99.tar.gz |

| Sequencing Project Name | File Name |
|---|---|
| LMW ETR NZ 2018 Waikite 6 1b MG | 3300045256_10.tar.gz |
| LMW ETR NZ 2018 Waikite 6 1b MG | 3300045256_100.tar.gz |
| LMW ETR NZ 2018 Waikite 6 1b MG | 3300045256_102.tar.gz |
| LMW ETR NZ 2018 Waikite 6 1b MG | 3300045256_103.tar.gz |
| LMW ETR NZ 2018 Waikite 6 1b MG | 3300045256_106.tar.gz |
| LMW ETR NZ 2018 Waikite 6 1b MG | 3300045256_109.tar.gz |
| LMW ETR NZ 2018 Waikite 6 1b MG | 3300045256_11.tar.gz |
| LMW ETR NZ 2018 Waikite 6 1b MG | 3300045256_111.tar.gz |
| LMW ETR NZ 2018 Waikite 6 1b MG | 3300045256_112.tar.gz |
| LMW ETR NZ 2018 Waikite 6 1b MG | 3300045256_114.tar.gz |
| LMW ETR NZ 2018 Waikite 6 1b MG | 3300045256_115.tar.gz |
| LMW ETR NZ 2018 Waikite 6 1b MG | 3300045256_117.tar.gz |
| LMW ETR NZ 2018 Waikite 6 1b MG | 3300045256_118.tar.gz |
| LMW ETR NZ 2018 Waikite 6 1b MG | 3300045256_119.tar.gz |
| LMW ETR NZ 2018 Waikite 6 1b MG | 3300045256_120.tar.gz |
| LMW ETR NZ 2018 Waikite 6 1b MG | 3300045256_121.tar.gz |
| LMW ETR NZ 2018 Waikite 6 1b MG | 3300045256_122.tar.gz |
| LMW ETR NZ 2018 Waikite 6 1b MG | 3300045256_123.tar.gz |
| LMW ETR NZ 2018 Waikite 6 1b MG | 3300045256_124.tar.gz |
| LMW ETR NZ 2018 Waikite 6 1b MG | 3300045256_126.tar.gz |
| LMW ETR NZ 2018 Waikite 6 1b MG | 3300045256_129.tar.gz |
| LMW ETR NZ 2018 Waikite 6 1b MG | 3300045256_13.tar.gz |
| LMW ETR NZ 2018 Waikite 6 1b MG | 3300045256_130.tar.gz |
| LMW ETR NZ 2018 Waikite 6 1b MG | 3300045256_131.tar.gz |
| LMW ETR NZ 2018 Waikite 6 1b MG | 3300045256_133.tar.gz |
| LMW ETR NZ 2018 Waikite 6 1b MG | 3300045256_137.tar.gz |
| LMW ETR NZ 2018 Waikite 6 1b MG | 3300045256_138.tar.gz |
| LMW ETR NZ 2018 Waikite 6 1b MG | 3300045256_139.tar.gz |
| LMW ETR NZ 2018 Waikite 6 1b MG | 3300045256_14.tar.gz |
| LMW ETR NZ 2018 Waikite 6 1b MG | 3300045256_140.tar.gz |
| LMW ETR NZ 2018 Waikite 6 1b MG | 3300045256_141.tar.gz |
| LMW ETR NZ 2018 Waikite 6 1b MG | 3300045256_146.tar.gz |
| LMW ETR NZ 2018 Waikite 6 1b MG | 3300045256_149.tar.gz |
| LMW ETR NZ 2018 Waikite 6 1b MG | 3300045256_15.tar.gz |
| LMW ETR NZ 2018 Waikite 6 1b MG | 3300045256_151.tar.gz |
| LMW ETR NZ 2018 Waikite 6 1b MG | 3300045256_152.tar.gz |
| LMW ETR NZ 2018 Waikite 6 1b MG | 3300045256_153.tar.gz |
| LMW ETR NZ 2018 Waikite 6 1b MG | 3300045256_154.tar.gz |
| LMW ETR NZ 2018 Waikite 6 1b MG | 3300045256_156.tar.gz |
| LMW ETR NZ 2018 Waikite 6 1b MG | 3300045256_157.tar.gz |
| LMW ETR NZ 2018 Waikite 6 1b MG | 3300045256_159.tar.gz |
| LMW ETR NZ 2018 Waikite 6 1b MG | 3300045256_161.tar.gz |
| LMW ETR NZ 2018 Waikite 6 1b MG | 3300045256_162.tar.gz |
| LMW ETR NZ 2018 Waikite 6 1b MG | 3300045256_165.tar.gz |
| LMW ETR NZ 2018 Waikite 6 1b MG | 3300045256_17.tar.gz |
| LMW ETR NZ 2018 Waikite 6 1b MG | 3300045256_170.tar.gz |
| LMW ETR NZ 2018 Waikite 6 1b MG | 3300045256_171.tar.gz |
| LMW ETR NZ 2018 Waikite 6 1b MG | 3300045256_173.tar.gz |
| LMW ETR NZ 2018 Waikite 6 1b MG | 3300045256_175.tar.gz |
| LMW ETR NZ 2018 Waikite 6 1b MG | 3300045256_176.tar.gz |
| LMW ETR NZ 2018 Waikite 6 1b MG | 3300045256_177.tar.gz |
| LMW ETR NZ 2018 Waikite 6 1b MG | 3300045256_18.tar.gz |
| LMW ETR NZ 2018 Waikite 6 1b MG | 3300045256_180.tar.gz |
| LMW ETR NZ 2018 Waikite 6 1b MG | 3300045256_182.tar.gz |
| LMW ETR NZ 2018 Waikite 6 1b MG | 3300045256_19.tar.gz |
| LMW ETR NZ 2018 Waikite 6 1b MG | 3300045256_2.tar.gz |
| LMW ETR NZ 2018 Waikite 6 1b MG | 3300045256_20.tar.gz |
| LMW ETR NZ 2018 Waikite 6 1b MG | 3300045256_21.tar.gz |
| LMW ETR NZ 2018 Waikite 6 1b MG | 3300045256_22.tar.gz |
| LMW ETR NZ 2018 Waikite 6 1b MG | 3300045256_23.tar.gz |
| LMW ETR NZ 2018 Waikite 6 1b MG | 3300045256_24.tar.gz |
| LMW ETR NZ 2018 Waikite 6 1b MG | 3300045256_25.tar.gz |
| LMW ETR NZ 2018 Waikite 6 1b MG | 3300045256_26.tar.gz |
| LMW ETR NZ 2018 Waikite 6 1b MG | 3300045256_27.tar.gz |
| LMW ETR NZ 2018 Waikite 6 1b MG | 3300045256_28.tar.gz |
| LMW ETR NZ 2018 Waikite 6 1b MG | 3300045256_33.tar.gz |
| LMW ETR NZ 2018 Waikite 6 1b MG | 3300045256_35.tar.gz |
| LMW ETR NZ 2018 Waikite 6 1b MG | 3300045256_39.tar.gz |
| LMW ETR NZ 2018 Waikite 6 1b MG | 3300045256_40.tar.gz |
| LMW ETR NZ 2018 Waikite 6 1b MG | 3300045256_41.tar.gz |
| LMW ETR NZ 2018 Waikite 6 1b MG | 3300045256_45.tar.gz |
| LMW ETR NZ 2018 Waikite 6 1b MG | 3300045256_49.tar.gz |
| LMW ETR NZ 2018 Waikite 6 1b MG | 3300045256_5.tar.gz |
| LMW ETR NZ 2018 Waikite 6 1b MG | 3300045256_51.tar.gz |
| LMW ETR NZ 2018 Waikite 6 1b MG | 3300045256_52.tar.gz |
| LMW ETR NZ 2018 Waikite 6 1b MG | 3300045256_53.tar.gz |
| LMW ETR NZ 2018 Waikite 6 1b MG | 3300045256_54.tar.gz |
| LMW ETR NZ 2018 Waikite 6 1b MG | 3300045256_55.tar.gz |
| LMW ETR NZ 2018 Waikite 6 1b MG | 3300045256_56.tar.gz |
| LMW ETR NZ 2018 Waikite 6 1b MG | 3300045256_59.tar.gz |
| LMW ETR NZ 2018 Waikite 6 1b MG | 3300045256_60.tar.gz |
| LMW ETR NZ 2018 Waikite 6 1b MG | 3300045256_65.tar.gz |
| LMW ETR NZ 2018 Waikite 6 1b MG | 3300045256_68.tar.gz |
| LMW ETR NZ 2018 Waikite 6 1b MG | 3300045256_69.tar.gz |
| LMW ETR NZ 2018 Waikite 6 1b MG | 3300045256_7.tar.gz |
| LMW ETR NZ 2018 Waikite 6 1b MG | 3300045256_70.tar.gz |
| LMW ETR NZ 2018 Waikite 6 1b MG | 3300045256_71.tar.gz |
| LMW ETR NZ 2018 Waikite 6 1b MG | 3300045256_72.tar.gz |
| LMW ETR NZ 2018 Waikite 6 1b MG | 3300045256_74.tar.gz |
| LMW ETR NZ 2018 Waikite 6 1b MG | 3300045256_75.tar.gz |
| LMW ETR NZ 2018 Waikite 6 1b MG | 3300045256_76.tar.gz |
| LMW ETR NZ 2018 Waikite 6 1b MG | 3300045256_77.tar.gz |
| LMW ETR NZ 2018 Waikite 6 1b MG | 3300045256_79.tar.gz |
| LMW ETR NZ 2018 Waikite 6 1b MG | 3300045256_80.tar.gz |
| LMW ETR NZ 2018 Waikite 6 1b MG | 3300045256_82.tar.gz |
| LMW ETR NZ 2018 Waikite 6 1b MG | 3300045256_83.tar.gz |
| LMW ETR NZ 2018 Waikite 6 1b MG | 3300045256_84.tar.gz |
| LMW ETR NZ 2018 Waikite 6 1b MG | 3300045256_85.tar.gz |
| LMW ETR NZ 2018 Waikite 6 1b MG | 3300045256_86.tar.gz |
| LMW ETR NZ 2018 Waikite 6 1b MG | 3300045256_87.tar.gz |
| LMW ETR NZ 2018 Waikite 6 1b MG | 3300045256_89.tar.gz |
| LMW ETR NZ 2018 Waikite 6 1b MG | 3300045256_91.tar.gz |
| LMW ETR NZ 2018 Waikite 6 1b MG | 3300045256_94.tar.gz |
| LMW ETR NZ 2018 Waikite 6 1b MG | 3300045256_97.tar.gz |

| Sequencing Project Name | File Name |
|---|---|
| LMW ETR NZ 2018 Waikite 6 1a MG | 3300045107_1.tar.gz |
| LMW ETR NZ 2018 Waikite 6 1a MG | 3300045107_10.tar.gz |
| LMW ETR NZ 2018 Waikite 6 1a MG | 3300045107_100.tar.gz |
| LMW ETR NZ 2018 Waikite 6 1a MG | 3300045107_102.tar.gz |
| LMW ETR NZ 2018 Waikite 6 1a MG | 3300045107_104.tar.gz |
| LMW ETR NZ 2018 Waikite 6 1a MG | 3300045107_105.tar.gz |
| LMW ETR NZ 2018 Waikite 6 1a MG | 3300045107_106.tar.gz |
| LMW ETR NZ 2018 Waikite 6 1a MG | 3300045107_108.tar.gz |
| LMW ETR NZ 2018 Waikite 6 1a MG | 3300045107_11.tar.gz |
| LMW ETR NZ 2018 Waikite 6 1a MG | 3300045107_110.tar.gz |
| LMW ETR NZ 2018 Waikite 6 1a MG | 3300045107_113.tar.gz |
| LMW ETR NZ 2018 Waikite 6 1a MG | 3300045107_114.tar.gz |
| LMW ETR NZ 2018 Waikite 6 1a MG | 3300045107_115.tar.gz |
| LMW ETR NZ 2018 Waikite 6 1a MG | 3300045107_120.tar.gz |
| LMW ETR NZ 2018 Waikite 6 1a MG | 3300045107_121.tar.gz |
| LMW ETR NZ 2018 Waikite 6 1a MG | 3300045107_122.tar.gz |
| LMW ETR NZ 2018 Waikite 6 1a MG | 3300045107_124.tar.gz |
| LMW ETR NZ 2018 Waikite 6 1a MG | 3300045107_125.tar.gz |
| LMW ETR NZ 2018 Waikite 6 1a MG | 3300045107_128.tar.gz |
| LMW ETR NZ 2018 Waikite 6 1a MG | 3300045107_129.tar.gz |
| LMW ETR NZ 2018 Waikite 6 1a MG | 3300045107_130.tar.gz |
| LMW ETR NZ 2018 Waikite 6 1a MG | 3300045107_134.tar.gz |
| LMW ETR NZ 2018 Waikite 6 1a MG | 3300045107_137.tar.gz |
| LMW ETR NZ 2018 Waikite 6 1a MG | 3300045107_138.tar.gz |
| LMW ETR NZ 2018 Waikite 6 1a MG | 3300045107_14.tar.gz |
| LMW ETR NZ 2018 Waikite 6 1a MG | 3300045107_140.tar.gz |
| LMW ETR NZ 2018 Waikite 6 1a MG | 3300045107_141.tar.gz |
| LMW ETR NZ 2018 Waikite 6 1a MG | 3300045107_144.tar.gz |
| LMW ETR NZ 2018 Waikite 6 1a MG | 3300045107_145.tar.gz |
| LMW ETR NZ 2018 Waikite 6 1a MG | 3300045107_146.tar.gz |
| LMW ETR NZ 2018 Waikite 6 1a MG | 3300045107_147.tar.gz |
| LMW ETR NZ 2018 Waikite 6 1a MG | 3300045107_149.tar.gz |
| LMW ETR NZ 2018 Waikite 6 1a MG | 3300045107_15.tar.gz |
| LMW ETR NZ 2018 Waikite 6 1a MG | 3300045107_150.tar.gz |
| LMW ETR NZ 2018 Waikite 6 1a MG | 3300045107_152.tar.gz |
| LMW ETR NZ 2018 Waikite 6 1a MG | 3300045107_155.tar.gz |
| LMW ETR NZ 2018 Waikite 6 1a MG | 3300045107_156.tar.gz |
| LMW ETR NZ 2018 Waikite 6 1a MG | 3300045107_157.tar.gz |
| LMW ETR NZ 2018 Waikite 6 1a MG | 3300045107_158.tar.gz |
| LMW ETR NZ 2018 Waikite 6 1a MG | 3300045107_16.tar.gz |
| LMW ETR NZ 2018 Waikite 6 1a MG | 3300045107_164.tar.gz |
| LMW ETR NZ 2018 Waikite 6 1a MG | 3300045107_165.tar.gz |
| LMW ETR NZ 2018 Waikite 6 1a MG | 3300045107_166.tar.gz |
| LMW ETR NZ 2018 Waikite 6 1a MG | 3300045107_167.tar.gz |
| LMW ETR NZ 2018 Waikite 6 1a MG | 3300045107_168.tar.gz |
| LMW ETR NZ 2018 Waikite 6 1a MG | 3300045107_169.tar.gz |
| LMW ETR NZ 2018 Waikite 6 1a MG | 3300045107_170.tar.gz |
| LMW ETR NZ 2018 Waikite 6 1a MG | 3300045107_171.tar.gz |
| LMW ETR NZ 2018 Waikite 6 1a MG | 3300045107_173.tar.gz |
| LMW ETR NZ 2018 Waikite 6 1a MG | 3300045107_174.tar.gz |
| LMW ETR NZ 2018 Waikite 6 1a MG | 3300045107_175.tar.gz |
| LMW ETR NZ 2018 Waikite 6 1a MG | 3300045107_179.tar.gz |
| LMW ETR NZ 2018 Waikite 6 1a MG | 3300045107_18.tar.gz |
| LMW ETR NZ 2018 Waikite 6 1a MG | 3300045107_181.tar.gz |
| LMW ETR NZ 2018 Waikite 6 1a MG | 3300045107_182.tar.gz |
| LMW ETR NZ 2018 Waikite 6 1a MG | 3300045107_2.tar.gz |
| LMW ETR NZ 2018 Waikite 6 1a MG | 3300045107_21.tar.gz |
| LMW ETR NZ 2018 Waikite 6 1a MG | 3300045107_24.tar.gz |
| LMW ETR NZ 2018 Waikite 6 1a MG | 3300045107_25.tar.gz |
| LMW ETR NZ 2018 Waikite 6 1a MG | 3300045107_26.tar.gz |
| LMW ETR NZ 2018 Waikite 6 1a MG | 3300045107_27.tar.gz |
| LMW ETR NZ 2018 Waikite 6 1a MG | 3300045107_28.tar.gz |
| LMW ETR NZ 2018 Waikite 6 1a MG | 3300045107_29.tar.gz |
| LMW ETR NZ 2018 Waikite 6 1a MG | 3300045107_3.tar.gz |
| LMW ETR NZ 2018 Waikite 6 1a MG | 3300045107_31.tar.gz |
| LMW ETR NZ 2018 Waikite 6 1a MG | 3300045107_32.tar.gz |
| LMW ETR NZ 2018 Waikite 6 1a MG | 3300045107_33.tar.gz |
| LMW ETR NZ 2018 Waikite 6 1a MG | 3300045107_37.tar.gz |
| LMW ETR NZ 2018 Waikite 6 1a MG | 3300045107_38.tar.gz |
| LMW ETR NZ 2018 Waikite 6 1a MG | 3300045107_39.tar.gz |
| LMW ETR NZ 2018 Waikite 6 1a MG | 3300045107_40.tar.gz |
| LMW ETR NZ 2018 Waikite 6 1a MG | 3300045107_41.tar.gz |
| LMW ETR NZ 2018 Waikite 6 1a MG | 3300045107_43.tar.gz |
| LMW ETR NZ 2018 Waikite 6 1a MG | 3300045107_46.tar.gz |
| LMW ETR NZ 2018 Waikite 6 1a MG | 3300045107_47.tar.gz |
| LMW ETR NZ 2018 Waikite 6 1a MG | 3300045107_48.tar.gz |
| LMW ETR NZ 2018 Waikite 6 1a MG | 3300045107_49.tar.gz |
| LMW ETR NZ 2018 Waikite 6 1a MG | 3300045107_5.tar.gz |
| LMW ETR NZ 2018 Waikite 6 1a MG | 3300045107_51.tar.gz |
| LMW ETR NZ 2018 Waikite 6 1a MG | 3300045107_52.tar.gz |
| LMW ETR NZ 2018 Waikite 6 1a MG | 3300045107_53.tar.gz |
| LMW ETR NZ 2018 Waikite 6 1a MG | 3300045107_55.tar.gz |
| LMW ETR NZ 2018 Waikite 6 1a MG | 3300045107_58.tar.gz |
| LMW ETR NZ 2018 Waikite 6 1a MG | 3300045107_60.tar.gz |
| LMW ETR NZ 2018 Waikite 6 1a MG | 3300045107_62.tar.gz |
| LMW ETR NZ 2018 Waikite 6 1a MG | 3300045107_63.tar.gz |
| LMW ETR NZ 2018 Waikite 6 1a MG | 3300045107_64.tar.gz |
| LMW ETR NZ 2018 Waikite 6 1a MG | 3300045107_65.tar.gz |
| LMW ETR NZ 2018 Waikite 6 1a MG | 3300045107_66.tar.gz |
| LMW ETR NZ 2018 Waikite 6 1a MG | 3300045107_67.tar.gz |
| LMW ETR NZ 2018 Waikite 6 1a MG | 3300045107_7.tar.gz |
| LMW ETR NZ 2018 Waikite 6 1a MG | 3300045107_70.tar.gz |
| LMW ETR NZ 2018 Waikite 6 1a MG | 3300045107_72.tar.gz |
| LMW ETR NZ 2018 Waikite 6 1a MG | 3300045107_73.tar.gz |

| LMW ETR NZ 2018 Waikite 6 1a MG | 3300045107_75.tar.gz |
| LMW ETR NZ 2018 Waikite 6 1a MG | 3300045107_76.tar.gz |
| LMW ETR NZ 2018 Waikite 6 1a MG | 3300045107_78.tar.gz |
| LMW ETR NZ 2018 Waikite 6 1a MG | 3300045107_79.tar.gz |
| LMW ETR NZ 2018 Waikite 6 1a MG | 3300045107_81.tar.gz |
| LMW ETR NZ 2018 Waikite 6 1a MG | 3300045107_84.tar.gz |
| LMW ETR NZ 2018 Waikite 6 1a MG | 3300045107_85.tar.gz |
| LMW ETR NZ 2018 Waikite 6 1a MG | 3300045107_87.tar.gz |
| LMW ETR NZ 2018 Waikite 6 1a MG | 3300045107_89.tar.gz |
| LMW ETR NZ 2018 Waikite 6 1a MG | 3300045107_9.tar.gz |
| LMW ETR NZ 2018 Waikite 6 1a MG | 3300045107_92.tar.gz |
| LMW ETR NZ 2018 Waikite 6 1a MG | 3300045107_93.tar.gz |
| LMW ETR NZ 2018 Waikite 6 1a MG | 3300045107_94.tar.gz |
| LMW ETR NZ 2018 Waikite 6 1a MG | 3300045107_96.tar.gz |
| LMW ETR NZ 2018 Waikite 6 1a MG | 3300045107_98.tar.gz |
| LMW ETR NZ 2018 Waikite 6 1a MG | 3300045107_99.tar.gz |

| Sequencing Project Name | File Name |
| --- | --- |
| LMW ETR NZ 2018 Waikite 5 2 MG | 3300044959_1.tar.gz |
| LMW ETR NZ 2018 Waikite 5 2 MG | 3300044959_17.tar.gz |
| LMW ETR NZ 2018 Waikite 5 2 MG | 3300044959_19.tar.gz |
| LMW ETR NZ 2018 Waikite 5 2 MG | 3300044959_20.tar.gz |
| LMW ETR NZ 2018 Waikite 5 2 MG | 3300044959_24.tar.gz |
| LMW ETR NZ 2018 Waikite 5 2 MG | 3300044959_25.tar.gz |
| LMW ETR NZ 2018 Waikite 5 2 MG | 3300044959_28.tar.gz |
| LMW ETR NZ 2018 Waikite 5 2 MG | 3300044959_29.tar.gz |
| LMW ETR NZ 2018 Waikite 5 2 MG | 3300044959_3.tar.gz |
| LMW ETR NZ 2018 Waikite 5 2 MG | 3300044959_33.tar.gz |
| LMW ETR NZ 2018 Waikite 5 2 MG | 3300044959_34.tar.gz |
| LMW ETR NZ 2018 Waikite 5 2 MG | 3300044959_35.tar.gz |
| LMW ETR NZ 2018 Waikite 5 2 MG | 3300044959_36.tar.gz |
| LMW ETR NZ 2018 Waikite 5 2 MG | 3300044959_37.tar.gz |
| LMW ETR NZ 2018 Waikite 5 2 MG | 3300044959_38.tar.gz |
| LMW ETR NZ 2018 Waikite 5 2 MG | 3300044959_39.tar.gz |
| LMW ETR NZ 2018 Waikite 5 2 MG | 3300044959_40.tar.gz |
| LMW ETR NZ 2018 Waikite 5 2 MG | 3300044959_41.tar.gz |
| LMW ETR NZ 2018 Waikite 5 2 MG | 3300044959_42.tar.gz |
| LMW ETR NZ 2018 Waikite 5 2 MG | 3300044959_47.tar.gz |
| LMW ETR NZ 2018 Waikite 5 2 MG | 3300044959_48.tar.gz |
| LMW ETR NZ 2018 Waikite 5 2 MG | 3300044959_49.tar.gz |
| LMW ETR NZ 2018 Waikite 5 2 MG | 3300044959_5.tar.gz |
| LMW ETR NZ 2018 Waikite 5 2 MG | 3300044959_50.tar.gz |
| LMW ETR NZ 2018 Waikite 5 2 MG | 3300044959_51.tar.gz |
| LMW ETR NZ 2018 Waikite 5 2 MG | 3300044959_53.tar.gz |
| LMW ETR NZ 2018 Waikite 5 2 MG | 3300044959_57.tar.gz |
| LMW ETR NZ 2018 Waikite 5 2 MG | 3300044959_60.tar.gz |
| LMW ETR NZ 2018 Waikite 5 2 MG | 3300044959_62.tar.gz |
| LMW ETR NZ 2018 Waikite 5 2 MG | 3300044959_64.tar.gz |
| LMW ETR NZ 2018 Waikite 5 2 MG | 3300044959_65.tar.gz |
| LMW ETR NZ 2018 Waikite 5 2 MG | 3300044959_66.tar.gz |
| LMW ETR NZ 2018 Waikite 5 2 MG | 3300044959_67.tar.gz |
| LMW ETR NZ 2018 Waikite 5 2 MG | 3300044959_69.tar.gz |
| LMW ETR NZ 2018 Waikite 5 2 MG | 3300044959_70.tar.gz |
| LMW ETR NZ 2018 Waikite 5 2 MG | 3300044959_72.tar.gz |
| LMW ETR NZ 2018 Waikite 5 2 MG | 3300044959_74.tar.gz |
| LMW ETR NZ 2018 Waikite 5 2 MG | 3300044959_77.tar.gz |
| LMW ETR NZ 2018 Waikite 5 2 MG | 3300044959_78.tar.gz |
| LMW ETR NZ 2018 Waikite 5 2 MG | 3300044959_8.tar.gz |
| LMW ETR NZ 2018 Waikite 5 2 MG | 3300044959_82.tar.gz |
| LMW ETR NZ 2018 Waikite 5 2 MG | 3300044959_83.tar.gz |
| LMW ETR NZ 2018 Waikite 5 2 MG | 3300044959_84.tar.gz |
| LMW ETR NZ 2018 Waikite 5 2 MG | 3300044959_85.tar.gz |
| LMW ETR NZ 2018 Waikite 5 2 MG | 3300044959_86.tar.gz |
| LMW ETR NZ 2018 Waikite 5 2 MG | 3300044959_87.tar.gz |
| LMW ETR NZ 2018 Waikite 5 2 MG | 3300044959_89.tar.gz |
| LMW ETR NZ 2018 Waikite 5 2 MG | 3300044959_9.tar.gz |

| Sequencing Project Name | File Name |
| --- | --- |
| LMW ETR NZ 2018 Waikite 5 1 MG | 3300045922_1.tar.gz |
| LMW ETR NZ 2018 Waikite 5 1 MG | 3300045922_101.tar.gz |
| LMW ETR NZ 2018 Waikite 5 1 MG | 3300045922_102.tar.gz |
| LMW ETR NZ 2018 Waikite 5 1 MG | 3300045922_103.tar.gz |
| LMW ETR NZ 2018 Waikite 5 1 MG | 3300045922_104.tar.gz |
| LMW ETR NZ 2018 Waikite 5 1 MG | 3300045922_105.tar.gz |
| LMW ETR NZ 2018 Waikite 5 1 MG | 3300045922_106.tar.gz |
| LMW ETR NZ 2018 Waikite 5 1 MG | 3300045922_112.tar.gz |
| LMW ETR NZ 2018 Waikite 5 1 MG | 3300045922_115.tar.gz |
| LMW ETR NZ 2018 Waikite 5 1 MG | 3300045922_118.tar.gz |
| LMW ETR NZ 2018 Waikite 5 1 MG | 3300045922_119.tar.gz |
| LMW ETR NZ 2018 Waikite 5 1 MG | 3300045922_12.tar.gz |
| LMW ETR NZ 2018 Waikite 5 1 MG | 3300045922_123.tar.gz |
| LMW ETR NZ 2018 Waikite 5 1 MG | 3300045922_127.tar.gz |
| LMW ETR NZ 2018 Waikite 5 1 MG | 3300045922_128.tar.gz |
| LMW ETR NZ 2018 Waikite 5 1 MG | 3300045922_129.tar.gz |
| LMW ETR NZ 2018 Waikite 5 1 MG | 3300045922_130.tar.gz |
| LMW ETR NZ 2018 Waikite 5 1 MG | 3300045922_131.tar.gz |
| LMW ETR NZ 2018 Waikite 5 1 MG | 3300045922_132.tar.gz |
| LMW ETR NZ 2018 Waikite 5 1 MG | 3300045922_133.tar.gz |
| LMW ETR NZ 2018 Waikite 5 1 MG | 3300045922_134.tar.gz |
| LMW ETR NZ 2018 Waikite 5 1 MG | 3300045922_135.tar.gz |
| LMW ETR NZ 2018 Waikite 5 1 MG | 3300045922_136.tar.gz |
| LMW ETR NZ 2018 Waikite 5 1 MG | 3300045922_137.tar.gz |
| LMW ETR NZ 2018 Waikite 5 1 MG | 3300045922_138.tar.gz |
| LMW ETR NZ 2018 Waikite 5 1 MG | 3300045922_139.tar.gz |
| LMW ETR NZ 2018 Waikite 5 1 MG | 3300045922_14.tar.gz |
| LMW ETR NZ 2018 Waikite 5 1 MG | 3300045922_140.tar.gz |
| LMW ETR NZ 2018 Waikite 5 1 MG | 3300045922_141.tar.gz |
| LMW ETR NZ 2018 Waikite 5 1 MG | 3300045922_142.tar.gz |
| LMW ETR NZ 2018 Waikite 5 1 MG | 3300045922_143.tar.gz |
| LMW ETR NZ 2018 Waikite 5 1 MG | 3300045922_145.tar.gz |
| LMW ETR NZ 2018 Waikite 5 1 MG | 3300045922_147.tar.gz |
| LMW ETR NZ 2018 Waikite 5 1 MG | 3300045922_15.tar.gz |
| LMW ETR NZ 2018 Waikite 5 1 MG | 3300045922_16.tar.gz |
| LMW ETR NZ 2018 Waikite 5 1 MG | 3300045922_17.tar.gz |
| LMW ETR NZ 2018 Waikite 5 1 MG | 3300045922_18.tar.gz |
| LMW ETR NZ 2018 Waikite 5 1 MG | 3300045922_2.tar.gz |
| LMW ETR NZ 2018 Waikite 5 1 MG | 3300045922_25.tar.gz |
| LMW ETR NZ 2018 Waikite 5 1 MG | 3300045922_26.tar.gz |
| LMW ETR NZ 2018 Waikite 5 1 MG | 3300045922_3.tar.gz |
| LMW ETR NZ 2018 Waikite 5 1 MG | 3300045922_31.tar.gz |
| LMW ETR NZ 2018 Waikite 5 1 MG | 3300045922_33.tar.gz |
| LMW ETR NZ 2018 Waikite 5 1 MG | 3300045922_38.tar.gz |
| LMW ETR NZ 2018 Waikite 5 1 MG | 3300045922_40.tar.gz |
| LMW ETR NZ 2018 Waikite 5 1 MG | 3300045922_41.tar.gz |
| LMW ETR NZ 2018 Waikite 5 1 MG | 3300045922_44.tar.gz |
| LMW ETR NZ 2018 Waikite 5 1 MG | 3300045922_45.tar.gz |
| LMW ETR NZ 2018 Waikite 5 1 MG | 3300045922_46.tar.gz |
| LMW ETR NZ 2018 Waikite 5 1 MG | 3300045922_47.tar.gz |
| LMW ETR NZ 2018 Waikite 5 1 MG | 3300045922_49.tar.gz |
| LMW ETR NZ 2018 Waikite 5 1 MG | 3300045922_5.tar.gz |
| LMW ETR NZ 2018 Waikite 5 1 MG | 3300045922_52.tar.gz |
| LMW ETR NZ 2018 Waikite 5 1 MG | 3300045922_53.tar.gz |
| LMW ETR NZ 2018 Waikite 5 1 MG | 3300045922_56.tar.gz |
| LMW ETR NZ 2018 Waikite 5 1 MG | 3300045922_58.tar.gz |
| LMW ETR NZ 2018 Waikite 5 1 MG | 3300045922_59.tar.gz |
| LMW ETR NZ 2018 Waikite 5 1 MG | 3300045922_6.tar.gz |
| LMW ETR NZ 2018 Waikite 5 1 MG | 3300045922_61.tar.gz |
| LMW ETR NZ 2018 Waikite 5 1 MG | 3300045922_63.tar.gz |
| LMW ETR NZ 2018 Waikite 5 1 MG | 3300045922_65.tar.gz |
| LMW ETR NZ 2018 Waikite 5 1 MG | 3300045922_67.tar.gz |
| LMW ETR NZ 2018 Waikite 5 1 MG | 3300045922_68.tar.gz |
| LMW ETR NZ 2018 Waikite 5 1 MG | 3300045922_69.tar.gz |
| LMW ETR NZ 2018 Waikite 5 1 MG | 3300045922_7.tar.gz |
| LMW ETR NZ 2018 Waikite 5 1 MG | 3300045922_72.tar.gz |
| LMW ETR NZ 2018 Waikite 5 1 MG | 3300045922_73.tar.gz |
| LMW ETR NZ 2018 Waikite 5 1 MG | 3300045922_74.tar.gz |
| LMW ETR NZ 2018 Waikite 5 1 MG | 3300045922_75.tar.gz |
| LMW ETR NZ 2018 Waikite 5 1 MG | 3300045922_76.tar.gz |
| LMW ETR NZ 2018 Waikite 5 1 MG | 3300045922_77.tar.gz |
| LMW ETR NZ 2018 Waikite 5 1 MG | 3300045922_8.tar.gz |
| LMW ETR NZ 2018 Waikite 5 1 MG | 3300045922_81.tar.gz |
| LMW ETR NZ 2018 Waikite 5 1 MG | 3300045922_84.tar.gz |
| LMW ETR NZ 2018 Waikite 5 1 MG | 3300045922_85.tar.gz |
| LMW ETR NZ 2018 Waikite 5 1 MG | 3300045922_86.tar.gz |
| LMW ETR NZ 2018 Waikite 5 1 MG | 3300045922_88.tar.gz |
| LMW ETR NZ 2018 Waikite 5 1 MG | 3300045922_93.tar.gz |
| LMW ETR NZ 2018 Waikite 5 1 MG | 3300045922_95.tar.gz |
| LMW ETR NZ 2018 Waikite 5 1 MG | 3300045922_97.tar.gz |
| LMW ETR NZ 2018 Waikite 5 1 MG | 3300045922_99.tar.gz |

| Sequencing Project Name | File Name |
|---|---|
| LMW ETR NZ 2018 Waikite 4 2 MG | 3300044941_10.tar.gz |
| LMW ETR NZ 2018 Waikite 4 2 MG | 3300044941_103.tar.gz |
| LMW ETR NZ 2018 Waikite 4 2 MG | 3300044941_105.tar.gz |
| LMW ETR NZ 2018 Waikite 4 2 MG | 3300044941_106.tar.gz |
| LMW ETR NZ 2018 Waikite 4 2 MG | 3300044941_107.tar.gz |
| LMW ETR NZ 2018 Waikite 4 2 MG | 3300044941_108.tar.gz |
| LMW ETR NZ 2018 Waikite 4 2 MG | 3300044941_109.tar.gz |
| LMW ETR NZ 2018 Waikite 4 2 MG | 3300044941_110.tar.gz |
| LMW ETR NZ 2018 Waikite 4 2 MG | 3300044941_112.tar.gz |
| LMW ETR NZ 2018 Waikite 4 2 MG | 3300044941_113.tar.gz |
| LMW ETR NZ 2018 Waikite 4 2 MG | 3300044941_13.tar.gz |
| LMW ETR NZ 2018 Waikite 4 2 MG | 3300044941_14.tar.gz |
| LMW ETR NZ 2018 Waikite 4 2 MG | 3300044941_19.tar.gz |
| LMW ETR NZ 2018 Waikite 4 2 MG | 3300044941_20.tar.gz |
| LMW ETR NZ 2018 Waikite 4 2 MG | 3300044941_23.tar.gz |
| LMW ETR NZ 2018 Waikite 4 2 MG | 3300044941_26.tar.gz |
| LMW ETR NZ 2018 Waikite 4 2 MG | 3300044941_27.tar.gz |
| LMW ETR NZ 2018 Waikite 4 2 MG | 3300044941_28.tar.gz |
| LMW ETR NZ 2018 Waikite 4 2 MG | 3300044941_3.tar.gz |
| LMW ETR NZ 2018 Waikite 4 2 MG | 3300044941_30.tar.gz |
| LMW ETR NZ 2018 Waikite 4 2 MG | 3300044941_31.tar.gz |
| LMW ETR NZ 2018 Waikite 4 2 MG | 3300044941_32.tar.gz |
| LMW ETR NZ 2018 Waikite 4 2 MG | 3300044941_33.tar.gz |
| LMW ETR NZ 2018 Waikite 4 2 MG | 3300044941_34.tar.gz |
| LMW ETR NZ 2018 Waikite 4 2 MG | 3300044941_35.tar.gz |
| LMW ETR NZ 2018 Waikite 4 2 MG | 3300044941_36.tar.gz |
| LMW ETR NZ 2018 Waikite 4 2 MG | 3300044941_37.tar.gz |
| LMW ETR NZ 2018 Waikite 4 2 MG | 3300044941_38.tar.gz |
| LMW ETR NZ 2018 Waikite 4 2 MG | 3300044941_39.tar.gz |
| LMW ETR NZ 2018 Waikite 4 2 MG | 3300044941_40.tar.gz |
| LMW ETR NZ 2018 Waikite 4 2 MG | 3300044941_41.tar.gz |
| LMW ETR NZ 2018 Waikite 4 2 MG | 3300044941_43.tar.gz |
| LMW ETR NZ 2018 Waikite 4 2 MG | 3300044941_44.tar.gz |
| LMW ETR NZ 2018 Waikite 4 2 MG | 3300044941_45.tar.gz |
| LMW ETR NZ 2018 Waikite 4 2 MG | 3300044941_46.tar.gz |
| LMW ETR NZ 2018 Waikite 4 2 MG | 3300044941_48.tar.gz |
| LMW ETR NZ 2018 Waikite 4 2 MG | 3300044941_49.tar.gz |
| LMW ETR NZ 2018 Waikite 4 2 MG | 3300044941_50.tar.gz |
| LMW ETR NZ 2018 Waikite 4 2 MG | 3300044941_51.tar.gz |
| LMW ETR NZ 2018 Waikite 4 2 MG | 3300044941_52.tar.gz |
| LMW ETR NZ 2018 Waikite 4 2 MG | 3300044941_54.tar.gz |
| LMW ETR NZ 2018 Waikite 4 2 MG | 3300044941_55.tar.gz |
| LMW ETR NZ 2018 Waikite 4 2 MG | 3300044941_56.tar.gz |
| LMW ETR NZ 2018 Waikite 4 2 MG | 3300044941_57.tar.gz |
| LMW ETR NZ 2018 Waikite 4 2 MG | 3300044941_59.tar.gz |
| LMW ETR NZ 2018 Waikite 4 2 MG | 3300044941_60.tar.gz |
| LMW ETR NZ 2018 Waikite 4 2 MG | 3300044941_61.tar.gz |
| LMW ETR NZ 2018 Waikite 4 2 MG | 3300044941_62.tar.gz |
| LMW ETR NZ 2018 Waikite 4 2 MG | 3300044941_63.tar.gz |
| LMW ETR NZ 2018 Waikite 4 2 MG | 3300044941_65.tar.gz |
| LMW ETR NZ 2018 Waikite 4 2 MG | 3300044941_66.tar.gz |
| LMW ETR NZ 2018 Waikite 4 2 MG | 3300044941_68.tar.gz |
| LMW ETR NZ 2018 Waikite 4 2 MG | 3300044941_69.tar.gz |
| LMW ETR NZ 2018 Waikite 4 2 MG | 3300044941_7.tar.gz |

| LMW ETR NZ 2018 Waikite 4 2 MG | 3300044941_70.tar.gz |
|---|---|
| LMW ETR NZ 2018 Waikite 4 2 MG | 3300044941_71.tar.gz |
| LMW ETR NZ 2018 Waikite 4 2 MG | 3300044941_72.tar.gz |
| LMW ETR NZ 2018 Waikite 4 2 MG | 3300044941_76.tar.gz |
| LMW ETR NZ 2018 Waikite 4 2 MG | 3300044941_82.tar.gz |
| LMW ETR NZ 2018 Waikite 4 2 MG | 3300044941_84.tar.gz |
| LMW ETR NZ 2018 Waikite 4 2 MG | 3300044941_85.tar.gz |
| LMW ETR NZ 2018 Waikite 4 2 MG | 3300044941_86.tar.gz |
| LMW ETR NZ 2018 Waikite 4 2 MG | 3300044941_9.tar.gz |
| LMW ETR NZ 2018 Waikite 4 2 MG | 3300044941_90.tar.gz |
| LMW ETR NZ 2018 Waikite 4 2 MG | 3300044941_91.tar.gz |
| LMW ETR NZ 2018 Waikite 4 2 MG | 3300044941_92.tar.gz |
| LMW ETR NZ 2018 Waikite 4 2 MG | 3300044941_94.tar.gz |
| LMW ETR NZ 2018 Waikite 4 2 MG | 3300044941_95.tar.gz |
| LMW ETR NZ 2018 Waikite 4 2 MG | 3300044941_96.tar.gz |
| LMW ETR NZ 2018 Waikite 4 2 MG | 3300044941_99.tar.gz |

| Sequencing Project Name | File Name |
|---|---|
| LMW ETR NZ 2018 Waikite 4 1 MG | 3300044940_1.tar.gz |
| LMW ETR NZ 2018 Waikite 4 1 MG | 3300044940_100.tar.gz |
| LMW ETR NZ 2018 Waikite 4 1 MG | 3300044940_101.tar.gz |
| LMW ETR NZ 2018 Waikite 4 1 MG | 3300044940_104.tar.gz |
| LMW ETR NZ 2018 Waikite 4 1 MG | 3300044940_107.tar.gz |
| LMW ETR NZ 2018 Waikite 4 1 MG | 3300044940_11.tar.gz |
| LMW ETR NZ 2018 Waikite 4 1 MG | 3300044940_110.tar.gz |
| LMW ETR NZ 2018 Waikite 4 1 MG | 3300044940_113.tar.gz |
| LMW ETR NZ 2018 Waikite 4 1 MG | 3300044940_116.tar.gz |
| LMW ETR NZ 2018 Waikite 4 1 MG | 3300044940_117.tar.gz |
| LMW ETR NZ 2018 Waikite 4 1 MG | 3300044940_12.tar.gz |
| LMW ETR NZ 2018 Waikite 4 1 MG | 3300044940_120.tar.gz |
| LMW ETR NZ 2018 Waikite 4 1 MG | 3300044940_121.tar.gz |
| LMW ETR NZ 2018 Waikite 4 1 MG | 3300044940_122.tar.gz |
| LMW ETR NZ 2018 Waikite 4 1 MG | 3300044940_125.tar.gz |
| LMW ETR NZ 2018 Waikite 4 1 MG | 3300044940_127.tar.gz |
| LMW ETR NZ 2018 Waikite 4 1 MG | 3300044940_13.tar.gz |
| LMW ETR NZ 2018 Waikite 4 1 MG | 3300044940_132.tar.gz |
| LMW ETR NZ 2018 Waikite 4 1 MG | 3300044940_133.tar.gz |
| LMW ETR NZ 2018 Waikite 4 1 MG | 3300044940_135.tar.gz |
| LMW ETR NZ 2018 Waikite 4 1 MG | 3300044940_136.tar.gz |
| LMW ETR NZ 2018 Waikite 4 1 MG | 3300044940_137.tar.gz |
| LMW ETR NZ 2018 Waikite 4 1 MG | 3300044940_139.tar.gz |
| LMW ETR NZ 2018 Waikite 4 1 MG | 3300044940_14.tar.gz |
| LMW ETR NZ 2018 Waikite 4 1 MG | 3300044940_140.tar.gz |
| LMW ETR NZ 2018 Waikite 4 1 MG | 3300044940_141.tar.gz |
| LMW ETR NZ 2018 Waikite 4 1 MG | 3300044940_144.tar.gz |
| LMW ETR NZ 2018 Waikite 4 1 MG | 3300044940_147.tar.gz |
| LMW ETR NZ 2018 Waikite 4 1 MG | 3300044940_148.tar.gz |
| LMW ETR NZ 2018 Waikite 4 1 MG | 3300044940_149.tar.gz |
| LMW ETR NZ 2018 Waikite 4 1 MG | 3300044940_18.tar.gz |
| LMW ETR NZ 2018 Waikite 4 1 MG | 3300044940_19.tar.gz |
| LMW ETR NZ 2018 Waikite 4 1 MG | 3300044940_22.tar.gz |
| LMW ETR NZ 2018 Waikite 4 1 MG | 3300044940_23.tar.gz |
| LMW ETR NZ 2018 Waikite 4 1 MG | 3300044940_24.tar.gz |
| LMW ETR NZ 2018 Waikite 4 1 MG | 3300044940_25.tar.gz |
| LMW ETR NZ 2018 Waikite 4 1 MG | 3300044940_26.tar.gz |
| LMW ETR NZ 2018 Waikite 4 1 MG | 3300044940_29.tar.gz |
| LMW ETR NZ 2018 Waikite 4 1 MG | 3300044940_3.tar.gz |
| LMW ETR NZ 2018 Waikite 4 1 MG | 3300044940_34.tar.gz |
| LMW ETR NZ 2018 Waikite 4 1 MG | 3300044940_35.tar.gz |
| LMW ETR NZ 2018 Waikite 4 1 MG | 3300044940_38.tar.gz |
| LMW ETR NZ 2018 Waikite 4 1 MG | 3300044940_39.tar.gz |
| LMW ETR NZ 2018 Waikite 4 1 MG | 3300044940_4.tar.gz |
| LMW ETR NZ 2018 Waikite 4 1 MG | 3300044940_40.tar.gz |
| LMW ETR NZ 2018 Waikite 4 1 MG | 3300044940_42.tar.gz |
| LMW ETR NZ 2018 Waikite 4 1 MG | 3300044940_44.tar.gz |
| LMW ETR NZ 2018 Waikite 4 1 MG | 3300044940_46.tar.gz |
| LMW ETR NZ 2018 Waikite 4 1 MG | 3300044940_48.tar.gz |
| LMW ETR NZ 2018 Waikite 4 1 MG | 3300044940_49.tar.gz |
| LMW ETR NZ 2018 Waikite 4 1 MG | 3300044940_5.tar.gz |
| LMW ETR NZ 2018 Waikite 4 1 MG | 3300044940_50.tar.gz |
| LMW ETR NZ 2018 Waikite 4 1 MG | 3300044940_51.tar.gz |
| LMW ETR NZ 2018 Waikite 4 1 MG | 3300044940_52.tar.gz |
| LMW ETR NZ 2018 Waikite 4 1 MG | 3300044940_53.tar.gz |
| LMW ETR NZ 2018 Waikite 4 1 MG | 3300044940_54.tar.gz |
| LMW ETR NZ 2018 Waikite 4 1 MG | 3300044940_55.tar.gz |
| LMW ETR NZ 2018 Waikite 4 1 MG | 3300044940_56.tar.gz |
| LMW ETR NZ 2018 Waikite 4 1 MG | 3300044940_57.tar.gz |
| LMW ETR NZ 2018 Waikite 4 1 MG | 3300044940_58.tar.gz |
| LMW ETR NZ 2018 Waikite 4 1 MG | 3300044940_59.tar.gz |
| LMW ETR NZ 2018 Waikite 4 1 MG | 3300044940_61.tar.gz |
| LMW ETR NZ 2018 Waikite 4 1 MG | 3300044940_63.tar.gz |
| LMW ETR NZ 2018 Waikite 4 1 MG | 3300044940_66.tar.gz |
| LMW ETR NZ 2018 Waikite 4 1 MG | 3300044940_67.tar.gz |
| LMW ETR NZ 2018 Waikite 4 1 MG | 3300044940_69.tar.gz |
| LMW ETR NZ 2018 Waikite 4 1 MG | 3300044940_7.tar.gz |
| LMW ETR NZ 2018 Waikite 4 1 MG | 3300044940_70.tar.gz |
| LMW ETR NZ 2018 Waikite 4 1 MG | 3300044940_71.tar.gz |
| LMW ETR NZ 2018 Waikite 4 1 MG | 3300044940_73.tar.gz |
| LMW ETR NZ 2018 Waikite 4 1 MG | 3300044940_75.tar.gz |
| LMW ETR NZ 2018 Waikite 4 1 MG | 3300044940_78.tar.gz |
| LMW ETR NZ 2018 Waikite 4 1 MG | 3300044940_79.tar.gz |
| LMW ETR NZ 2018 Waikite 4 1 MG | 3300044940_80.tar.gz |
| LMW ETR NZ 2018 Waikite 4 1 MG | 3300044940_81.tar.gz |
| LMW ETR NZ 2018 Waikite 4 1 MG | 3300044940_82.tar.gz |
| LMW ETR NZ 2018 Waikite 4 1 MG | 3300044940_84.tar.gz |
| LMW ETR NZ 2018 Waikite 4 1 MG | 3300044940_85.tar.gz |
| LMW ETR NZ 2018 Waikite 4 1 MG | 3300044940_88.tar.gz |
| LMW ETR NZ 2018 Waikite 4 1 MG | 3300044940_89.tar.gz |
| LMW ETR NZ 2018 Waikite 4 1 MG | 3300044940_9.tar.gz |
| LMW ETR NZ 2018 Waikite 4 1 MG | 3300044940_91.tar.gz |
| LMW ETR NZ 2018 Waikite 4 1 MG | 3300044940_92.tar.gz |
| LMW ETR NZ 2018 Waikite 4 1 MG | 3300044940_94.tar.gz |
| LMW ETR NZ 2018 Waikite 4 1 MG | 3300044940_95.tar.gz |

| Sequencing Project Name | File Name |
|---|---|
| LMW ETR NZ 2018 Waikite 4 1 MG | 3300044940_98.tar.gz |

| Sequencing Project Name | File Name |
|---|---|
| LMW ETR NZ 2018 Waikite 3 2 MG | 3300044939_10.tar.gz |
| LMW ETR NZ 2018 Waikite 3 2 MG | 3300044939_11.tar.gz |
| LMW ETR NZ 2018 Waikite 3 2 MG | 3300044939_12.tar.gz |
| LMW ETR NZ 2018 Waikite 3 2 MG | 3300044939_17.tar.gz |
| LMW ETR NZ 2018 Waikite 3 2 MG | 3300044939_18.tar.gz |
| LMW ETR NZ 2018 Waikite 3 2 MG | 3300044939_2.tar.gz |
| LMW ETR NZ 2018 Waikite 3 2 MG | 3300044939_20.tar.gz |
| LMW ETR NZ 2018 Waikite 3 2 MG | 3300044939_21.tar.gz |
| LMW ETR NZ 2018 Waikite 3 2 MG | 3300044939_24.tar.gz |
| LMW ETR NZ 2018 Waikite 3 2 MG | 3300044939_26.tar.gz |
| LMW ETR NZ 2018 Waikite 3 2 MG | 3300044939_27.tar.gz |
| LMW ETR NZ 2018 Waikite 3 2 MG | 3300044939_3.tar.gz |
| LMW ETR NZ 2018 Waikite 3 2 MG | 3300044939_30.tar.gz |
| LMW ETR NZ 2018 Waikite 3 2 MG | 3300044939_31.tar.gz |
| LMW ETR NZ 2018 Waikite 3 2 MG | 3300044939_33.tar.gz |
| LMW ETR NZ 2018 Waikite 3 2 MG | 3300044939_36.tar.gz |
| LMW ETR NZ 2018 Waikite 3 2 MG | 3300044939_37.tar.gz |
| LMW ETR NZ 2018 Waikite 3 2 MG | 3300044939_38.tar.gz |
| LMW ETR NZ 2018 Waikite 3 2 MG | 3300044939_39.tar.gz |
| LMW ETR NZ 2018 Waikite 3 2 MG | 3300044939_4.tar.gz |
| LMW ETR NZ 2018 Waikite 3 2 MG | 3300044939_40.tar.gz |
| LMW ETR NZ 2018 Waikite 3 2 MG | 3300044939_42.tar.gz |
| LMW ETR NZ 2018 Waikite 3 2 MG | 3300044939_44.tar.gz |
| LMW ETR NZ 2018 Waikite 3 2 MG | 3300044939_46.tar.gz |
| LMW ETR NZ 2018 Waikite 3 2 MG | 3300044939_48.tar.gz |
| LMW ETR NZ 2018 Waikite 3 2 MG | 3300044939_49.tar.gz |
| LMW ETR NZ 2018 Waikite 3 2 MG | 3300044939_51.tar.gz |
| LMW ETR NZ 2018 Waikite 3 2 MG | 3300044939_52.tar.gz |
| LMW ETR NZ 2018 Waikite 3 2 MG | 3300044939_55.tar.gz |
| LMW ETR NZ 2018 Waikite 3 2 MG | 3300044939_56.tar.gz |

| LMW ETR NZ 2018 Waikite 3 2 MG | 3300044939_57.tar.gz |
| LMW ETR NZ 2018 Waikite 3 2 MG | 3300044939_58.tar.gz |
| LMW ETR NZ 2018 Waikite 3 2 MG | 3300044939_59.tar.gz |
| LMW ETR NZ 2018 Waikite 3 2 MG | 3300044939_6.tar.gz |
| LMW ETR NZ 2018 Waikite 3 2 MG | 3300044939_61.tar.gz |
| LMW ETR NZ 2018 Waikite 3 2 MG | 3300044939_66.tar.gz |
| LMW ETR NZ 2018 Waikite 3 2 MG | 3300044939_67.tar.gz |
| LMW ETR NZ 2018 Waikite 3 2 MG | 3300044939_68.tar.gz |
| LMW ETR NZ 2018 Waikite 3 2 MG | 3300044939_69.tar.gz |
| LMW ETR NZ 2018 Waikite 3 2 MG | 3300044939_7.tar.gz |
| LMW ETR NZ 2018 Waikite 3 2 MG | 3300044939_70.tar.gz |
| LMW ETR NZ 2018 Waikite 3 2 MG | 3300044939_72.tar.gz |
| LMW ETR NZ 2018 Waikite 3 2 MG | 3300044939_79.tar.gz |
| LMW ETR NZ 2018 Waikite 3 2 MG | 3300044939_80.tar.gz |
| LMW ETR NZ 2018 Waikite 3 2 MG | 3300044939_81.tar.gz |
| LMW ETR NZ 2018 Waikite 3 2 MG | 3300044939_83.tar.gz |
| LMW ETR NZ 2018 Waikite 3 2 MG | 3300044939_85.tar.gz |
| LMW ETR NZ 2018 Waikite 3 2 MG | 3300044939_87.tar.gz |
| LMW ETR NZ 2018 Waikite 3 2 MG | 3300044939_88.tar.gz |
| LMW ETR NZ 2018 Waikite 3 2 MG | 3300044939_9.tar.gz |
| LMW ETR NZ 2018 Waikite 3 2 MG | 3300044939_91.tar.gz |

| Sequencing Project Name | File Name |
|---|---|
| LMW ETR NZ 2018 Waikite 3 1 MG | 3300044995_14.tar.gz |
| LMW ETR NZ 2018 Waikite 3 1 MG | 3300044995_21.tar.gz |
| LMW ETR NZ 2018 Waikite 3 1 MG | 3300044995_22.tar.gz |
| LMW ETR NZ 2018 Waikite 3 1 MG | 3300044995_24.tar.gz |
| LMW ETR NZ 2018 Waikite 3 1 MG | 3300044995_26.tar.gz |
| LMW ETR NZ 2018 Waikite 3 1 MG | 3300044995_27.tar.gz |
| LMW ETR NZ 2018 Waikite 3 1 MG | 3300044995_28.tar.gz |
| LMW ETR NZ 2018 Waikite 3 1 MG | 3300044995_30.tar.gz |
| LMW ETR NZ 2018 Waikite 3 1 MG | 3300044995_31.tar.gz |
| LMW ETR NZ 2018 Waikite 3 1 MG | 3300044995_34.tar.gz |
| LMW ETR NZ 2018 Waikite 3 1 MG | 3300044995_36.tar.gz |
| LMW ETR NZ 2018 Waikite 3 1 MG | 3300044995_37.tar.gz |
| LMW ETR NZ 2018 Waikite 3 1 MG | 3300044995_38.tar.gz |
| LMW ETR NZ 2018 Waikite 3 1 MG | 3300044995_39.tar.gz |
| LMW ETR NZ 2018 Waikite 3 1 MG | 3300044995_4.tar.gz |
| LMW ETR NZ 2018 Waikite 3 1 MG | 3300044995_40.tar.gz |
| LMW ETR NZ 2018 Waikite 3 1 MG | 3300044995_42.tar.gz |
| LMW ETR NZ 2018 Waikite 3 1 MG | 3300044995_44.tar.gz |
| LMW ETR NZ 2018 Waikite 3 1 MG | 3300044995_45.tar.gz |
| LMW ETR NZ 2018 Waikite 3 1 MG | 3300044995_46.tar.gz |
| LMW ETR NZ 2018 Waikite 3 1 MG | 3300044995_5.tar.gz |
| LMW ETR NZ 2018 Waikite 3 1 MG | 3300044995_50.tar.gz |
| LMW ETR NZ 2018 Waikite 3 1 MG | 3300044995_51.tar.gz |
| LMW ETR NZ 2018 Waikite 3 1 MG | 3300044995_53.tar.gz |
| LMW ETR NZ 2018 Waikite 3 1 MG | 3300044995_54.tar.gz |
| LMW ETR NZ 2018 Waikite 3 1 MG | 3300044995_55.tar.gz |
| LMW ETR NZ 2018 Waikite 3 1 MG | 3300044995_56.tar.gz |
| LMW ETR NZ 2018 Waikite 3 1 MG | 3300044995_58.tar.gz |
| LMW ETR NZ 2018 Waikite 3 1 MG | 3300044995_59.tar.gz |
| LMW ETR NZ 2018 Waikite 3 1 MG | 3300044995_60.tar.gz |
| LMW ETR NZ 2018 Waikite 3 1 MG | 3300044995_63.tar.gz |
| LMW ETR NZ 2018 Waikite 3 1 MG | 3300044995_67.tar.gz |
| LMW ETR NZ 2018 Waikite 3 1 MG | 3300044995_68.tar.gz |
| LMW ETR NZ 2018 Waikite 3 1 MG | 3300044995_69.tar.gz |
| LMW ETR NZ 2018 Waikite 3 1 MG | 3300044995_75.tar.gz |
| LMW ETR NZ 2018 Waikite 3 1 MG | 3300044995_76.tar.gz |
| LMW ETR NZ 2018 Waikite 3 1 MG | 3300044995_77.tar.gz |
| LMW ETR NZ 2018 Waikite 3 1 MG | 3300044995_78.tar.gz |
| LMW ETR NZ 2018 Waikite 3 1 MG | 3300044995_79.tar.gz |
| LMW ETR NZ 2018 Waikite 3 1 MG | 3300044995_81.tar.gz |
| LMW ETR NZ 2018 Waikite 3 1 MG | 3300044995_85.tar.gz |
| LMW ETR NZ 2018 Waikite 3 1 MG | 3300044995_88.tar.gz |
| LMW ETR NZ 2018 Waikite 3 1 MG | 3300044995_89.tar.gz |
| LMW ETR NZ 2018 Waikite 3 1 MG | 3300044995_9.tar.gz |
| LMW ETR NZ 2018 Waikite 3 1 MG | 3300044995_90.tar.gz |

| Sequencing Project Name | File Name |
| --- | --- |
| LMW ETR NZ 2018 Waikite 3 1 MG | 3300044995_92.tar.gz |
| LMW ETR NZ 2018 Waikite 3 1 MG | 3300044995_94.tar.gz |

| Sequencing Project Name | File Name |
| --- | --- |
| LMW ETR NZ 2018 Waikite 2 2 MG | 3300045921_10.tar.gz |
| LMW ETR NZ 2018 Waikite 2 2 MG | 3300045921_100.tar.gz |
| LMW ETR NZ 2018 Waikite 2 2 MG | 3300045921_101.tar.gz |
| LMW ETR NZ 2018 Waikite 2 2 MG | 3300045921_106.tar.gz |
| LMW ETR NZ 2018 Waikite 2 2 MG | 3300045921_110.tar.gz |
| LMW ETR NZ 2018 Waikite 2 2 MG | 3300045921_112.tar.gz |
| LMW ETR NZ 2018 Waikite 2 2 MG | 3300045921_113.tar.gz |
| LMW ETR NZ 2018 Waikite 2 2 MG | 3300045921_114.tar.gz |
| LMW ETR NZ 2018 Waikite 2 2 MG | 3300045921_116.tar.gz |
| LMW ETR NZ 2018 Waikite 2 2 MG | 3300045921_118.tar.gz |
| LMW ETR NZ 2018 Waikite 2 2 MG | 3300045921_120.tar.gz |
| LMW ETR NZ 2018 Waikite 2 2 MG | 3300045921_129.tar.gz |
| LMW ETR NZ 2018 Waikite 2 2 MG | 3300045921_13.tar.gz |
| LMW ETR NZ 2018 Waikite 2 2 MG | 3300045921_130.tar.gz |
| LMW ETR NZ 2018 Waikite 2 2 MG | 3300045921_133.tar.gz |
| LMW ETR NZ 2018 Waikite 2 2 MG | 3300045921_134.tar.gz |
| LMW ETR NZ 2018 Waikite 2 2 MG | 3300045921_135.tar.gz |
| LMW ETR NZ 2018 Waikite 2 2 MG | 3300045921_138.tar.gz |
| LMW ETR NZ 2018 Waikite 2 2 MG | 3300045921_14.tar.gz |
| LMW ETR NZ 2018 Waikite 2 2 MG | 3300045921_142.tar.gz |
| LMW ETR NZ 2018 Waikite 2 2 MG | 3300045921_143.tar.gz |
| LMW ETR NZ 2018 Waikite 2 2 MG | 3300045921_147.tar.gz |
| LMW ETR NZ 2018 Waikite 2 2 MG | 3300045921_148.tar.gz |
| LMW ETR NZ 2018 Waikite 2 2 MG | 3300045921_150.tar.gz |
| LMW ETR NZ 2018 Waikite 2 2 MG | 3300045921_152.tar.gz |
| LMW ETR NZ 2018 Waikite 2 2 MG | 3300045921_154.tar.gz |
| LMW ETR NZ 2018 Waikite 2 2 MG | 3300045921_17.tar.gz |
| LMW ETR NZ 2018 Waikite 2 2 MG | 3300045921_18.tar.gz |
| LMW ETR NZ 2018 Waikite 2 2 MG | 3300045921_19.tar.gz |
| LMW ETR NZ 2018 Waikite 2 2 MG | 3300045921_2.tar.gz |
| LMW ETR NZ 2018 Waikite 2 2 MG | 3300045921_23.tar.gz |
| LMW ETR NZ 2018 Waikite 2 2 MG | 3300045921_24.tar.gz |
| LMW ETR NZ 2018 Waikite 2 2 MG | 3300045921_25.tar.gz |
| LMW ETR NZ 2018 Waikite 2 2 MG | 3300045921_26.tar.gz |
| LMW ETR NZ 2018 Waikite 2 2 MG | 3300045921_28.tar.gz |
| LMW ETR NZ 2018 Waikite 2 2 MG | 3300045921_3.tar.gz |
| LMW ETR NZ 2018 Waikite 2 2 MG | 3300045921_31.tar.gz |
| LMW ETR NZ 2018 Waikite 2 2 MG | 3300045921_32.tar.gz |
| LMW ETR NZ 2018 Waikite 2 2 MG | 3300045921_33.tar.gz |
| LMW ETR NZ 2018 Waikite 2 2 MG | 3300045921_34.tar.gz |
| LMW ETR NZ 2018 Waikite 2 2 MG | 3300045921_35.tar.gz |
| LMW ETR NZ 2018 Waikite 2 2 MG | 3300045921_36.tar.gz |
| LMW ETR NZ 2018 Waikite 2 2 MG | 3300045921_38.tar.gz |
| LMW ETR NZ 2018 Waikite 2 2 MG | 3300045921_40.tar.gz |
| LMW ETR NZ 2018 Waikite 2 2 MG | 3300045921_41.tar.gz |
| LMW ETR NZ 2018 Waikite 2 2 MG | 3300045921_42.tar.gz |
| LMW ETR NZ 2018 Waikite 2 2 MG | 3300045921_43.tar.gz |
| LMW ETR NZ 2018 Waikite 2 2 MG | 3300045921_46.tar.gz |
| LMW ETR NZ 2018 Waikite 2 2 MG | 3300045921_47.tar.gz |
| LMW ETR NZ 2018 Waikite 2 2 MG | 3300045921_48.tar.gz |
| LMW ETR NZ 2018 Waikite 2 2 MG | 3300045921_49.tar.gz |
| LMW ETR NZ 2018 Waikite 2 2 MG | 3300045921_50.tar.gz |
| LMW ETR NZ 2018 Waikite 2 2 MG | 3300045921_52.tar.gz |
| LMW ETR NZ 2018 Waikite 2 2 MG | 3300045921_53.tar.gz |
| LMW ETR NZ 2018 Waikite 2 2 MG | 3300045921_56.tar.gz |
| LMW ETR NZ 2018 Waikite 2 2 MG | 3300045921_57.tar.gz |
| LMW ETR NZ 2018 Waikite 2 2 MG | 3300045921_58.tar.gz |
| LMW ETR NZ 2018 Waikite 2 2 MG | 3300045921_59.tar.gz |
| LMW ETR NZ 2018 Waikite 2 2 MG | 3300045921_60.tar.gz |
| LMW ETR NZ 2018 Waikite 2 2 MG | 3300045921_61.tar.gz |
| LMW ETR NZ 2018 Waikite 2 2 MG | 3300045921_62.tar.gz |
| LMW ETR NZ 2018 Waikite 2 2 MG | 3300045921_63.tar.gz |
| LMW ETR NZ 2018 Waikite 2 2 MG | 3300045921_64.tar.gz |
| LMW ETR NZ 2018 Waikite 2 2 MG | 3300045921_65.tar.gz |
| LMW ETR NZ 2018 Waikite 2 2 MG | 3300045921_66.tar.gz |

| LMW ETR NZ 2018 Waikite 2 2 MG | 3300045921_67.tar.gz |
|---|---|
| LMW ETR NZ 2018 Waikite 2 2 MG | 3300045921_68.tar.gz |
| LMW ETR NZ 2018 Waikite 2 2 MG | 3300045921_69.tar.gz |
| LMW ETR NZ 2018 Waikite 2 2 MG | 3300045921_7.tar.gz |
| LMW ETR NZ 2018 Waikite 2 2 MG | 3300045921_70.tar.gz |
| LMW ETR NZ 2018 Waikite 2 2 MG | 3300045921_72.tar.gz |
| LMW ETR NZ 2018 Waikite 2 2 MG | 3300045921_73.tar.gz |
| LMW ETR NZ 2018 Waikite 2 2 MG | 3300045921_75.tar.gz |
| LMW ETR NZ 2018 Waikite 2 2 MG | 3300045921_78.tar.gz |
| LMW ETR NZ 2018 Waikite 2 2 MG | 3300045921_79.tar.gz |
| LMW ETR NZ 2018 Waikite 2 2 MG | 3300045921_81.tar.gz |
| LMW ETR NZ 2018 Waikite 2 2 MG | 3300045921_86.tar.gz |
| LMW ETR NZ 2018 Waikite 2 2 MG | 3300045921_88.tar.gz |
| LMW ETR NZ 2018 Waikite 2 2 MG | 3300045921_89.tar.gz |
| LMW ETR NZ 2018 Waikite 2 2 MG | 3300045921_91.tar.gz |
| LMW ETR NZ 2018 Waikite 2 2 MG | 3300045921_92.tar.gz |
| LMW ETR NZ 2018 Waikite 2 2 MG | 3300045921_97.tar.gz |
| LMW ETR NZ 2018 Waikite 2 2 MG | 3300045921_98.tar.gz |

| Sequencing Project Name | File Name |
|---|---|
| LMW ETR NZ 2018 Waikite 2 1 MG | 3300045911_1.tar.gz |
| LMW ETR NZ 2018 Waikite 2 1 MG | 3300045911_102.tar.gz |
| LMW ETR NZ 2018 Waikite 2 1 MG | 3300045911_105.tar.gz |
| LMW ETR NZ 2018 Waikite 2 1 MG | 3300045911_106.tar.gz |
| LMW ETR NZ 2018 Waikite 2 1 MG | 3300045911_107.tar.gz |
| LMW ETR NZ 2018 Waikite 2 1 MG | 3300045911_113.tar.gz |
| LMW ETR NZ 2018 Waikite 2 1 MG | 3300045911_114.tar.gz |
| LMW ETR NZ 2018 Waikite 2 1 MG | 3300045911_116.tar.gz |
| LMW ETR NZ 2018 Waikite 2 1 MG | 3300045911_117.tar.gz |
| LMW ETR NZ 2018 Waikite 2 1 MG | 3300045911_12.tar.gz |
| LMW ETR NZ 2018 Waikite 2 1 MG | 3300045911_120.tar.gz |
| LMW ETR NZ 2018 Waikite 2 1 MG | 3300045911_122.tar.gz |
| LMW ETR NZ 2018 Waikite 2 1 MG | 3300045911_123.tar.gz |
| LMW ETR NZ 2018 Waikite 2 1 MG | 3300045911_124.tar.gz |
| LMW ETR NZ 2018 Waikite 2 1 MG | 3300045911_128.tar.gz |
| LMW ETR NZ 2018 Waikite 2 1 MG | 3300045911_130.tar.gz |
| LMW ETR NZ 2018 Waikite 2 1 MG | 3300045911_131.tar.gz |
| LMW ETR NZ 2018 Waikite 2 1 MG | 3300045911_132.tar.gz |
| LMW ETR NZ 2018 Waikite 2 1 MG | 3300045911_133.tar.gz |
| LMW ETR NZ 2018 Waikite 2 1 MG | 3300045911_134.tar.gz |
| LMW ETR NZ 2018 Waikite 2 1 MG | 3300045911_135.tar.gz |
| LMW ETR NZ 2018 Waikite 2 1 MG | 3300045911_136.tar.gz |
| LMW ETR NZ 2018 Waikite 2 1 MG | 3300045911_137.tar.gz |
| LMW ETR NZ 2018 Waikite 2 1 MG | 3300045911_139.tar.gz |
| LMW ETR NZ 2018 Waikite 2 1 MG | 3300045911_140.tar.gz |
| LMW ETR NZ 2018 Waikite 2 1 MG | 3300045911_141.tar.gz |
| LMW ETR NZ 2018 Waikite 2 1 MG | 3300045911_143.tar.gz |
| LMW ETR NZ 2018 Waikite 2 1 MG | 3300045911_144.tar.gz |
| LMW ETR NZ 2018 Waikite 2 1 MG | 3300045911_15.tar.gz |
| LMW ETR NZ 2018 Waikite 2 1 MG | 3300045911_150.tar.gz |
| LMW ETR NZ 2018 Waikite 2 1 MG | 3300045911_151.tar.gz |
| LMW ETR NZ 2018 Waikite 2 1 MG | 3300045911_153.tar.gz |
| LMW ETR NZ 2018 Waikite 2 1 MG | 3300045911_16.tar.gz |
| LMW ETR NZ 2018 Waikite 2 1 MG | 3300045911_17.tar.gz |
| LMW ETR NZ 2018 Waikite 2 1 MG | 3300045911_18.tar.gz |
| LMW ETR NZ 2018 Waikite 2 1 MG | 3300045911_2.tar.gz |
| LMW ETR NZ 2018 Waikite 2 1 MG | 3300045911_21.tar.gz |
| LMW ETR NZ 2018 Waikite 2 1 MG | 3300045911_22.tar.gz |
| LMW ETR NZ 2018 Waikite 2 1 MG | 3300045911_3.tar.gz |
| LMW ETR NZ 2018 Waikite 2 1 MG | 3300045911_32.tar.gz |
| LMW ETR NZ 2018 Waikite 2 1 MG | 3300045911_34.tar.gz |
| LMW ETR NZ 2018 Waikite 2 1 MG | 3300045911_36.tar.gz |
| LMW ETR NZ 2018 Waikite 2 1 MG | 3300045911_37.tar.gz |
| LMW ETR NZ 2018 Waikite 2 1 MG | 3300045911_38.tar.gz |
| LMW ETR NZ 2018 Waikite 2 1 MG | 3300045911_39.tar.gz |
| LMW ETR NZ 2018 Waikite 2 1 MG | 3300045911_4.tar.gz |
| LMW ETR NZ 2018 Waikite 2 1 MG | 3300045911_41.tar.gz |
| LMW ETR NZ 2018 Waikite 2 1 MG | 3300045911_44.tar.gz |
| LMW ETR NZ 2018 Waikite 2 1 MG | 3300045911_46.tar.gz |
| LMW ETR NZ 2018 Waikite 2 1 MG | 3300045911_47.tar.gz |
| LMW ETR NZ 2018 Waikite 2 1 MG | 3300045911_48.tar.gz |
| LMW ETR NZ 2018 Waikite 2 1 MG | 3300045911_49.tar.gz |
| LMW ETR NZ 2018 Waikite 2 1 MG | 3300045911_50.tar.gz |
| LMW ETR NZ 2018 Waikite 2 1 MG | 3300045911_51.tar.gz |
| LMW ETR NZ 2018 Waikite 2 1 MG | 3300045911_53.tar.gz |
| LMW ETR NZ 2018 Waikite 2 1 MG | 3300045911_58.tar.gz |
| LMW ETR NZ 2018 Waikite 2 1 MG | 3300045911_59.tar.gz |
| LMW ETR NZ 2018 Waikite 2 1 MG | 3300045911_60.tar.gz |
| LMW ETR NZ 2018 Waikite 2 1 MG | 3300045911_61.tar.gz |
| LMW ETR NZ 2018 Waikite 2 1 MG | 3300045911_64.tar.gz |
| LMW ETR NZ 2018 Waikite 2 1 MG | 3300045911_66.tar.gz |
| LMW ETR NZ 2018 Waikite 2 1 MG | 3300045911_67.tar.gz |
| LMW ETR NZ 2018 Waikite 2 1 MG | 3300045911_68.tar.gz |
| LMW ETR NZ 2018 Waikite 2 1 MG | 3300045911_69.tar.gz |
| LMW ETR NZ 2018 Waikite 2 1 MG | 3300045911_7.tar.gz |
| LMW ETR NZ 2018 Waikite 2 1 MG | 3300045911_71.tar.gz |
| LMW ETR NZ 2018 Waikite 2 1 MG | 3300045911_72.tar.gz |
| LMW ETR NZ 2018 Waikite 2 1 MG | 3300045911_73.tar.gz |
| LMW ETR NZ 2018 Waikite 2 1 MG | 3300045911_74.tar.gz |
| LMW ETR NZ 2018 Waikite 2 1 MG | 3300045911_75.tar.gz |
| LMW ETR NZ 2018 Waikite 2 1 MG | 3300045911_76.tar.gz |
| LMW ETR NZ 2018 Waikite 2 1 MG | 3300045911_77.tar.gz |
| LMW ETR NZ 2018 Waikite 2 1 MG | 3300045911_78.tar.gz |
| LMW ETR NZ 2018 Waikite 2 1 MG | 3300045911_79.tar.gz |
| LMW ETR NZ 2018 Waikite 2 1 MG | 3300045911_8.tar.gz |
| LMW ETR NZ 2018 Waikite 2 1 MG | 3300045911_80.tar.gz |
| LMW ETR NZ 2018 Waikite 2 1 MG | 3300045911_81.tar.gz |
| LMW ETR NZ 2018 Waikite 2 1 MG | 3300045911_83.tar.gz |
| LMW ETR NZ 2018 Waikite 2 1 MG | 3300045911_84.tar.gz |
| LMW ETR NZ 2018 Waikite 2 1 MG | 3300045911_86.tar.gz |
| LMW ETR NZ 2018 Waikite 2 1 MG | 3300045911_87.tar.gz |
| LMW ETR NZ 2018 Waikite 2 1 MG | 3300045911_89.tar.gz |
| LMW ETR NZ 2018 Waikite 2 1 MG | 3300045911_90.tar.gz |
| LMW ETR NZ 2018 Waikite 2 1 MG | 3300045911_91.tar.gz |
| LMW ETR NZ 2018 Waikite 2 1 MG | 3300045911_93.tar.gz |
| LMW ETR NZ 2018 Waikite 2 1 MG | 3300045911_94.tar.gz |
| LMW ETR NZ 2018 Waikite 2 1 MG | 3300045911_96.tar.gz |

| Sequencing Project Name | File Name |
|---|---|
| LMW ETR NZ 2018 Waikite 1 2 MG | 3300044994_1.tar.gz |
| LMW ETR NZ 2018 Waikite 1 2 MG | 3300044994_10.tar.gz |
| LMW ETR NZ 2018 Waikite 1 2 MG | 3300044994_101.tar.gz |
| LMW ETR NZ 2018 Waikite 1 2 MG | 3300044994_104.tar.gz |
| LMW ETR NZ 2018 Waikite 1 2 MG | 3300044994_106.tar.gz |
| LMW ETR NZ 2018 Waikite 1 2 MG | 3300044994_108.tar.gz |
| LMW ETR NZ 2018 Waikite 1 2 MG | 3300044994_109.tar.gz |
| LMW ETR NZ 2018 Waikite 1 2 MG | 3300044994_11.tar.gz |
| LMW ETR NZ 2018 Waikite 1 2 MG | 3300044994_110.tar.gz |
| LMW ETR NZ 2018 Waikite 1 2 MG | 3300044994_111.tar.gz |
| LMW ETR NZ 2018 Waikite 1 2 MG | 3300044994_112.tar.gz |
| LMW ETR NZ 2018 Waikite 1 2 MG | 3300044994_113.tar.gz |
| LMW ETR NZ 2018 Waikite 1 2 MG | 3300044994_114.tar.gz |
| LMW ETR NZ 2018 Waikite 1 2 MG | 3300044994_116.tar.gz |
| LMW ETR NZ 2018 Waikite 1 2 MG | 3300044994_117.tar.gz |
| LMW ETR NZ 2018 Waikite 1 2 MG | 3300044994_13.tar.gz |
| LMW ETR NZ 2018 Waikite 1 2 MG | 3300044994_18.tar.gz |
| LMW ETR NZ 2018 Waikite 1 2 MG | 3300044994_2.tar.gz |
| LMW ETR NZ 2018 Waikite 1 2 MG | 3300044994_20.tar.gz |
| LMW ETR NZ 2018 Waikite 1 2 MG | 3300044994_22.tar.gz |
| LMW ETR NZ 2018 Waikite 1 2 MG | 3300044994_23.tar.gz |
| LMW ETR NZ 2018 Waikite 1 2 MG | 3300044994_24.tar.gz |
| LMW ETR NZ 2018 Waikite 1 2 MG | 3300044994_26.tar.gz |
| LMW ETR NZ 2018 Waikite 1 2 MG | 3300044994_27.tar.gz |
| LMW ETR NZ 2018 Waikite 1 2 MG | 3300044994_31.tar.gz |
| LMW ETR NZ 2018 Waikite 1 2 MG | 3300044994_33.tar.gz |
| LMW ETR NZ 2018 Waikite 1 2 MG | 3300044994_34.tar.gz |
| LMW ETR NZ 2018 Waikite 1 2 MG | 3300044994_35.tar.gz |
| LMW ETR NZ 2018 Waikite 1 2 MG | 3300044994_36.tar.gz |
| LMW ETR NZ 2018 Waikite 1 2 MG | 3300044994_37.tar.gz |
| LMW ETR NZ 2018 Waikite 1 2 MG | 3300044994_38.tar.gz |
| LMW ETR NZ 2018 Waikite 1 2 MG | 3300044994_39.tar.gz |
| LMW ETR NZ 2018 Waikite 1 2 MG | 3300044994_4.tar.gz |
| LMW ETR NZ 2018 Waikite 1 2 MG | 3300044994_40.tar.gz |
| LMW ETR NZ 2018 Waikite 1 2 MG | 3300044994_42.tar.gz |
| LMW ETR NZ 2018 Waikite 1 2 MG | 3300044994_43.tar.gz |
| LMW ETR NZ 2018 Waikite 1 2 MG | 3300044994_44.tar.gz |
| LMW ETR NZ 2018 Waikite 1 2 MG | 3300044994_45.tar.gz |
| LMW ETR NZ 2018 Waikite 1 2 MG | 3300044994_46.tar.gz |
| LMW ETR NZ 2018 Waikite 1 2 MG | 3300044994_47.tar.gz |
| LMW ETR NZ 2018 Waikite 1 2 MG | 3300044994_5.tar.gz |
| LMW ETR NZ 2018 Waikite 1 2 MG | 3300044994_50.tar.gz |
| LMW ETR NZ 2018 Waikite 1 2 MG | 3300044994_52.tar.gz |
| LMW ETR NZ 2018 Waikite 1 2 MG | 3300044994_54.tar.gz |
| LMW ETR NZ 2018 Waikite 1 2 MG | 3300044994_55.tar.gz |
| LMW ETR NZ 2018 Waikite 1 2 MG | 3300044994_56.tar.gz |
| LMW ETR NZ 2018 Waikite 1 2 MG | 3300044994_57.tar.gz |
| LMW ETR NZ 2018 Waikite 1 2 MG | 3300044994_59.tar.gz |
| LMW ETR NZ 2018 Waikite 1 2 MG | 3300044994_61.tar.gz |
| LMW ETR NZ 2018 Waikite 1 2 MG | 3300044994_62.tar.gz |
| LMW ETR NZ 2018 Waikite 1 2 MG | 3300044994_64.tar.gz |
| LMW ETR NZ 2018 Waikite 1 2 MG | 3300044994_67.tar.gz |
| LMW ETR NZ 2018 Waikite 1 2 MG | 3300044994_68.tar.gz |
| LMW ETR NZ 2018 Waikite 1 2 MG | 3300044994_69.tar.gz |
| LMW ETR NZ 2018 Waikite 1 2 MG | 3300044994_7.tar.gz |
| LMW ETR NZ 2018 Waikite 1 2 MG | 3300044994_70.tar.gz |
| LMW ETR NZ 2018 Waikite 1 2 MG | 3300044994_71.tar.gz |
| LMW ETR NZ 2018 Waikite 1 2 MG | 3300044994_74.tar.gz |
| LMW ETR NZ 2018 Waikite 1 2 MG | 3300044994_75.tar.gz |
| LMW ETR NZ 2018 Waikite 1 2 MG | 3300044994_76.tar.gz |
| LMW ETR NZ 2018 Waikite 1 2 MG | 3300044994_79.tar.gz |
| LMW ETR NZ 2018 Waikite 1 2 MG | 3300044994_8.tar.gz |
| LMW ETR NZ 2018 Waikite 1 2 MG | 3300044994_82.tar.gz |
| LMW ETR NZ 2018 Waikite 1 2 MG | 3300044994_85.tar.gz |

| Sequencing Project Name | File Name |
| --- | --- |
| LMW ETR NZ 2018 Waikite 1 2 MG | 3300044994_87.tar.gz |
| LMW ETR NZ 2018 Waikite 1 2 MG | 3300044994_91.tar.gz |
| LMW ETR NZ 2018 Waikite 1 2 MG | 3300044994_96.tar.gz |
| LMW ETR NZ 2018 Waikite 1 2 MG | 3300044994_98.tar.gz |
| LMW ETR NZ 2018 Waikite 1 2 MG | 3300044994_99.tar.gz |

| Sequencing Project Name | File Name |
| --- | --- |
| LMW ETR NZ 2018 Waikite 1 1 MG | 3300044993_102.tar.gz |
| LMW ETR NZ 2018 Waikite 1 1 MG | 3300044993_103.tar.gz |
| LMW ETR NZ 2018 Waikite 1 1 MG | 3300044993_104.tar.gz |
| LMW ETR NZ 2018 Waikite 1 1 MG | 3300044993_106.tar.gz |
| LMW ETR NZ 2018 Waikite 1 1 MG | 3300044993_107.tar.gz |
| LMW ETR NZ 2018 Waikite 1 1 MG | 3300044993_109.tar.gz |
| LMW ETR NZ 2018 Waikite 1 1 MG | 3300044993_11.tar.gz |
| LMW ETR NZ 2018 Waikite 1 1 MG | 3300044993_111.tar.gz |
| LMW ETR NZ 2018 Waikite 1 1 MG | 3300044993_113.tar.gz |
| LMW ETR NZ 2018 Waikite 1 1 MG | 3300044993_117.tar.gz |
| LMW ETR NZ 2018 Waikite 1 1 MG | 3300044993_119.tar.gz |
| LMW ETR NZ 2018 Waikite 1 1 MG | 3300044993_120.tar.gz |
| LMW ETR NZ 2018 Waikite 1 1 MG | 3300044993_123.tar.gz |
| LMW ETR NZ 2018 Waikite 1 1 MG | 3300044993_129.tar.gz |
| LMW ETR NZ 2018 Waikite 1 1 MG | 3300044993_13.tar.gz |
| LMW ETR NZ 2018 Waikite 1 1 MG | 3300044993_131.tar.gz |
| LMW ETR NZ 2018 Waikite 1 1 MG | 3300044993_132.tar.gz |
| LMW ETR NZ 2018 Waikite 1 1 MG | 3300044993_133.tar.gz |
| LMW ETR NZ 2018 Waikite 1 1 MG | 3300044993_134.tar.gz |
| LMW ETR NZ 2018 Waikite 1 1 MG | 3300044993_135.tar.gz |
| LMW ETR NZ 2018 Waikite 1 1 MG | 3300044993_136.tar.gz |
| LMW ETR NZ 2018 Waikite 1 1 MG | 3300044993_137.tar.gz |
| LMW ETR NZ 2018 Waikite 1 1 MG | 3300044993_138.tar.gz |
| LMW ETR NZ 2018 Waikite 1 1 MG | 3300044993_139.tar.gz |
| LMW ETR NZ 2018 Waikite 1 1 MG | 3300044993_140.tar.gz |
| LMW ETR NZ 2018 Waikite 1 1 MG | 3300044993_141.tar.gz |
| LMW ETR NZ 2018 Waikite 1 1 MG | 3300044993_143.tar.gz |
| LMW ETR NZ 2018 Waikite 1 1 MG | 3300044993_146.tar.gz |
| LMW ETR NZ 2018 Waikite 1 1 MG | 3300044993_148.tar.gz |
| LMW ETR NZ 2018 Waikite 1 1 MG | 3300044993_15.tar.gz |
| LMW ETR NZ 2018 Waikite 1 1 MG | 3300044993_2.tar.gz |
| LMW ETR NZ 2018 Waikite 1 1 MG | 3300044993_20.tar.gz |
| LMW ETR NZ 2018 Waikite 1 1 MG | 3300044993_22.tar.gz |
| LMW ETR NZ 2018 Waikite 1 1 MG | 3300044993_23.tar.gz |
| LMW ETR NZ 2018 Waikite 1 1 MG | 3300044993_24.tar.gz |
| LMW ETR NZ 2018 Waikite 1 1 MG | 3300044993_25.tar.gz |
| LMW ETR NZ 2018 Waikite 1 1 MG | 3300044993_28.tar.gz |
| LMW ETR NZ 2018 Waikite 1 1 MG | 3300044993_29.tar.gz |
| LMW ETR NZ 2018 Waikite 1 1 MG | 3300044993_30.tar.gz |
| LMW ETR NZ 2018 Waikite 1 1 MG | 3300044993_33.tar.gz |
| LMW ETR NZ 2018 Waikite 1 1 MG | 3300044993_34.tar.gz |
| LMW ETR NZ 2018 Waikite 1 1 MG | 3300044993_43.tar.gz |
| LMW ETR NZ 2018 Waikite 1 1 MG | 3300044993_44.tar.gz |
| LMW ETR NZ 2018 Waikite 1 1 MG | 3300044993_45.tar.gz |
| LMW ETR NZ 2018 Waikite 1 1 MG | 3300044993_47.tar.gz |
| LMW ETR NZ 2018 Waikite 1 1 MG | 3300044993_49.tar.gz |
| LMW ETR NZ 2018 Waikite 1 1 MG | 3300044993_5.tar.gz |
| LMW ETR NZ 2018 Waikite 1 1 MG | 3300044993_50.tar.gz |
| LMW ETR NZ 2018 Waikite 1 1 MG | 3300044993_52.tar.gz |
| LMW ETR NZ 2018 Waikite 1 1 MG | 3300044993_53.tar.gz |
| LMW ETR NZ 2018 Waikite 1 1 MG | 3300044993_54.tar.gz |
| LMW ETR NZ 2018 Waikite 1 1 MG | 3300044993_55.tar.gz |
| LMW ETR NZ 2018 Waikite 1 1 MG | 3300044993_56.tar.gz |
| LMW ETR NZ 2018 Waikite 1 1 MG | 3300044993_57.tar.gz |
| LMW ETR NZ 2018 Waikite 1 1 MG | 3300044993_58.tar.gz |
| LMW ETR NZ 2018 Waikite 1 1 MG | 3300044993_59.tar.gz |
| LMW ETR NZ 2018 Waikite 1 1 MG | 3300044993_6.tar.gz |
| LMW ETR NZ 2018 Waikite 1 1 MG | 3300044993_61.tar.gz |
| LMW ETR NZ 2018 Waikite 1 1 MG | 3300044993_62.tar.gz |
| LMW ETR NZ 2018 Waikite 1 1 MG | 3300044993_64.tar.gz |
| LMW ETR NZ 2018 Waikite 1 1 MG | 3300044993_65.tar.gz |
| LMW ETR NZ 2018 Waikite 1 1 MG | 3300044993_66.tar.gz |
| LMW ETR NZ 2018 Waikite 1 1 MG | 3300044993_67.tar.gz |
| LMW ETR NZ 2018 Waikite 1 1 MG | 3300044993_68.tar.gz |
| LMW ETR NZ 2018 Waikite 1 1 MG | 3300044993_69.tar.gz |
| LMW ETR NZ 2018 Waikite 1 1 MG | 3300044993_71.tar.gz |
| LMW ETR NZ 2018 Waikite 1 1 MG | 3300044993_72.tar.gz |
| LMW ETR NZ 2018 Waikite 1 1 MG | 3300044993_73.tar.gz |
| LMW ETR NZ 2018 Waikite 1 1 MG | 3300044993_74.tar.gz |
| LMW ETR NZ 2018 Waikite 1 1 MG | 3300044993_75.tar.gz |
| LMW ETR NZ 2018 Waikite 1 1 MG | 3300044993_76.tar.gz |
| LMW ETR NZ 2018 Waikite 1 1 MG | 3300044993_77.tar.gz |
| LMW ETR NZ 2018 Waikite 1 1 MG | 3300044993_78.tar.gz |
| LMW ETR NZ 2018 Waikite 1 1 MG | 3300044993_8.tar.gz |
| LMW ETR NZ 2018 Waikite 1 1 MG | 3300044993_80.tar.gz |
| LMW ETR NZ 2018 Waikite 1 1 MG | 3300044993_81.tar.gz |
| LMW ETR NZ 2018 Waikite 1 1 MG | 3300044993_84.tar.gz |
| LMW ETR NZ 2018 Waikite 1 1 MG | 3300044993_86.tar.gz |
| LMW ETR NZ 2018 Waikite 1 1 MG | 3300044993_87.tar.gz |
| LMW ETR NZ 2018 Waikite 1 1 MG | 3300044993_88.tar.gz |
| LMW ETR NZ 2018 Waikite 1 1 MG | 3300044993_90.tar.gz |
| LMW ETR NZ 2018 Waikite 1 1 MG | 3300044993_91.tar.gz |
| LMW ETR NZ 2018 Waikite 1 1 MG | 3300044993_92.tar.gz |
| LMW ETR NZ 2018 Waikite 1 1 MG | 3300044993_94.tar.gz |
| LMW ETR NZ 2018 Waikite 1 1 MG | 3300044993_97.tar.gz |
| LMW ETR NZ 2018 Waikite 1 1 MG | 3300044993_98.tar.gz |
| LMW ETR NZ 2018 Waikite 1 1 MG | 3300044993_99.tar.gz |